\documentclass[a4paper,11pt]{article}

\usepackage{amsmath, amssymb, bm}
\usepackage[]{graphicx}
\usepackage[caption=false]{subfig}
\usepackage{fullpage}
\usepackage{authblk}
\usepackage{hyperref}
\newcommand{\be}{\begin{eqnarray}}
\newcommand{\ee}{\end{eqnarray}}

\begin{document}

\title{\bf{CMB constraints on mass and coupling constant of light pseudoscalar particles}}
\author[1,3]{Damian Ejlli}
\author[2,3,4]{Alexander D. Dolgov}
\affil[1]{\emph{Theory group, INFN Laboratori Nazionali del Gran Sasso, 67100 Assergi, Italy}}
\affil[2]{\emph{Dipartimento di Fisica e Scienze della Terra, Universita degli studi di Ferrara, 44122 Ferrara, Italy}}
\affil[3]{\emph{Department of Physics, Novosibirsk State University, 
Novosibirsk 630090, Russia}}
\affil[4]{\emph{ITEP, 113259 Moscow, Russia}}

\date{}

\maketitle

\begin{abstract}
Transformation of CMB photons into light pseudoscalar particles at post big bang nucleosynthesis epoch is considered. Using 
the present day value of a large scale magnetic field to estimate it at earlier cosmological epochs, the oscillation probability of photons into 
light pseudoscalar particles with an account of coherence breaking in cosmological plasma is calculated. Demanding that the  
photon transformation does not lead to an exceedingly large CMB spectral distortion and temperature anisotropy, 
the constraints on the coupling constant of axion like particles to photons,  
$ g_{\phi\gamma} B \lesssim (10^{-15} - 10^{-12}) \textrm{nG}\times \textrm{GeV}^{-1}$, 
are found  for the axion like particle mass in the interval $10^{-25}$ eV $\lesssim m_{\phi}\lesssim 10^{-5}$ eV, where $B$ is  the strength
of  the large scale magnetic field at the present time. Our results update the previously obtained ones since we use the density matrix
formalism which is more accurate than the wave function approximation for the description of oscillations with an essential coherence breaking. 
In the axion like particle mass range $10^{-25}$ eV $\lesssim m_{\phi}\lesssim 10^{-14}$ eV, weaker limits,  by at least 2 orders of magnitude $g_{\phi\gamma} B \lesssim 10^{-11}\textrm{nG}\times \textrm{GeV}^{-1}$,  are obtained in comparison with the wave function approximation. In the mass range  $10^{-14}$ eV $\lesssim m_{\phi}\lesssim 10^{-5}$ eV, on the other hand, limits that are stronger, by more than an order of magnitude are obtained. Our results are derived by using upper limits on spectral distortion parameter $\mu$ and temperature anisotropy $\Delta T/T$ found by COBE and expected sensitivities by PIXIE/PRISM.
\end{abstract}

\href{mailto:ejlli@fe.infn.it}{ejlli@fe.infn.it} \quad \href{mailto:dolgov@fe.infn.it}{dolgov@fe.infn.it}

\section{\label{sec:Intro}Introduction}

The mixing of particles with different spins in a magnetic field may produce interesting effects in physics and especially in cosmology. In particular, 
the mixing of axions with photons can be essential in some astrophysical and cosmological processes. 
Originally proposed as a solution to the strong CP problem \cite{Peccei:1977hh}, axions have been extensively studied in many contexts.
One of the most important implications of axions in cosmology is that they are good candidates for cold dark matter, which in particular,
could  form a Bose-Einstein condensate \cite{Sikivie:2009fv}. On the other hand, several extensions of the standard model and string 
theory~\cite{Arvanitaki:2009fg} also postulate an existence of light pseudo-scalar particles or axion like particles (ALPs) \cite{Ringwald:2012hr} which are good candidates of cold dark matter as well. Direct experimental search of ALPs are being performed by the CAST experiment at CERN, the PVLAS experiment in Italy, and at DESY in Germany. The light shining through a wall experiment or ALPs-I at DESY at the moment is being upgraded to ALPs-II which, by 2017, aims to find ALPs with a coupling constant to photons $10^{-11}\textrm{GeV}^{-1}\lesssim g_{\phi\gamma}$ for $m_\phi\lesssim 10^{-3}$ eV where $m_\phi$ is the ALP mass; see Ref. \cite{Bahre:2013ywa}.

There is an
essential difference between axions and ALPs in the relation between their masses and coupling constants to photons, $g_{a\gamma}, g_{\phi\gamma}$ (axions, ALPs). 
In the former case there is a linear connection between $g_{a\gamma}$ and the axion mass, $M_a$, while in the latter case there 
is no \emph{a priori} relationship between the ALP mass, $m_\phi$, and the coupling constant to photons, $g_{\phi\gamma}$. The ALP 
coupling to two photons may play an important role in cosmology because of the transformation of ALPs into photons or vice versa
in the presence of large scale magnetic fields. An analogous phenomenon of graviton-photon oscillations 
can be induced in an external magnetic due to similar coupling of a graviton to two photons.
However, there is an essential difference between the two cases which is related to the different values of the graviton and ALP masses
and their coupling constants to photons.
In the case of ALPs the two parameters are unrelated and 
\emph{not known}, except for some upper bounds, while in the case of gravitons the coupling constant is rigorously fixed by the General
Relativity (GR) and is very small, since it is inversely proportional to the Planck mass, $m_{Pl}$. In GR the graviton must be
exactly massless. In this paper we do not consider the case of gravitons. Oscillations between cosmic microwave background (CMB) photons into gravitons were considered in Ref. \cite{Ejlli:2013lsa}.

The probability of photon-ALP transformation depends on the strength of the large scale magnetic field, $B$, and on its coherence
length, $\lambda_B$. In particular, the mixing angle is proportional to  $B$. 
The propagation of ALPs or photons in the magnetic field is usually studied in two limiting cases.  
The first possibility is relatively short  $\lambda_B$ of the order of the intergalactic distance $\sim$ Mpc, so 
the photon to ALP transformation effectively takes place when 
they pass through several domains of the magnetic field with a random direction of the field vector $\mathbf B$. 
Under such conditions the transformation probability depends on the size of the domains (coherence length of the magnetic fields), 
their number, and on the momentum of ALPs or photons~\cite{Grossman:2002by}. In this case 
the typical upper limits on field strength at $\lambda_B\simeq 1$ Mpc are
$B\lesssim 6\times 10^{-8} - 2\times 10^{-6}$ G from the Faraday rotation of CMB and $B\lesssim 3\times 10^{-9}$ G from 
the CMB angular anisotropy, see Ref. \cite{Paoletti:2012bb}. 


The second possibility is that there is a large scale homogeneous magnetic field with a coherence length of the order of the
Hubble horizon at the present time. Consequently, ALPs or photons pass through just one magnetic domain. In this case the Faraday rotation of CMB, Lyman $\alpha$-forest, and temperature anisotropy put upper limits on the field strength of the order of 
$B\lesssim (1 - 4)\times 10^{-9}$ G for $\lambda_B$ comparable to the horizon scale, see Ref. \cite{Barrow:1997mj}. 
Since the existence of intergalactic magnetic fields and their spatial structure is not well known, there is no \emph{a priori} reason 
to assume either the first possibility or the second one. In this work we consider the second possibility and 
take as a fiducial value of the magnetic field today $B\lesssim 10^{-9}$ G and its direction as fixed in the sky. Moreover, 
we work under the commonly used assumption that magnetic field lines are frozen in the cosmological plasma (flux conservation)
and that the field strength scales with the cosmological scale factor, $a$, as $B=B_i(a_i/a)^2$.
The observational data on large scale magnetic fields are discussed in Ref.~\cite{magn-obs}
and a for review on the origin, strength, and conservation of the magnetic field flux see e.g. Ref. \cite{Grasso:2000wj}.

The early universe is a promising place to look for the transformation of CMB photons into ALPs 
or vice versa. This effect was first considered in Ref. \cite{Yanagida:1987nf} in the resonant case and the latter in Ref. \cite{Mirizzi:2009nq}, where more detailed calculations were done in the resonant case including a possible increase of the free electron number density during the reionization epoch. 
Using the upper limits on temperature anisotropy and spectral distortion of CMB obtained by the COBE/FIRAS instrument \cite{Fixsen:1996nj}, the authors of Ref. \cite{Mirizzi:2009nq} put the bound on the product 
$g_{\phi\gamma}B\lesssim 10^{-13}-10^{-11}$ GeV$^{-1}$nG for the 
ALP masses $m_\phi\simeq 10^{-14}-10^{-4}$ eV (see Fig. 3 of Ref. \cite{Mirizzi:2009nq}). The resonant case has been recently 
revised in Ref. \cite{Tashiro:2013yea} on the basis of the anticipated improvement on measurements of the CMB frequency 
spectrum by the recently proposed space missions PIXIE \cite{Kogut:2011xw} and PRISM \cite{Andre:2013afa}. 
They will improve the sensitivity at least by 3 orders of magnitude with respect to COBE and the authors 
of  Ref. \cite{Tashiro:2013yea} derived the expected constraint $g_{\phi\gamma}B\lesssim 10^{-16}$ GeV$^{-1}$nG 
for the ALPs masses $m_\phi\lesssim 10^{-14}$ eV.

In this work we aim to study the mixing of photons with ALPs and consider their oscillations  in the large scale magnetic field in resonant and 
non resonant cases with an account of the coherence breaking of the oscillations due to non-forward elastic and inelastic scattering
 in the cosmological plasma. In order to deal with this 
problem one needs to go beyond the simple wave function approximation and to solve the quantum kinetic 
equations for the density operator of the photon-ALP system. 
Such an approach allows us to include the scattering and absorption/annihilation of photons and ALPs in the medium and is applicable both to resonance and
non resonance cases. Consequently we update the existing bound on the ALP parameter space and extend it for the ALP masses below $10^{-14}$ eV. For the ALP masses $m_\phi <10^{-14}$ eV we find that density matrix formalism gives weaker results by at least 2 orders of magnitude with respect to the wave function approach. For ALP masses $m_\phi>10^{-14}$ eV we present new results which in most cases are stronger than those found by 
the wave function approach. 

The  paper is organized as follows. In Sec.~\ref{sec:2} we derive in the WKB approximation
the equations of motions for the axion-ALP system in a background magnetic field. In Sec.~\ref{sec:3} we calculate the essential decoherence effects in the photon-ALP system in the cosmological plasma; in Sec.~\ref{sec:4} we solve the equations of motion in the steady state approximation for the resonant case. In Sec.~\ref{sec:5} we numerically solve equations of motion in the general case, and in Sec.~\ref{sec:6} we conclude. 
In this paper we use the natural units, $c=\hbar=k_B=1$.

\section{\label{sec:2}Photon-pseudoscalar mixing in magnetic field. Equations of motion}

The total action describing the interactions of photons with ALPs is  given by the sum of the following terms:
\begin{align}\nonumber\label{em-action}
S=& -\frac{1}{4}\int d^4xF_{\mu\nu}F^{\mu\nu}+\frac{\alpha^2}{90m_e^4}\int d^4x\left[(F_{\mu\nu}F^{\mu\nu})^2+\frac{7}{4}(\tilde F_{\mu\nu}F^{\mu\nu})^2\right] + \\ &\frac{1}{2}\int d^4x\,(\partial_\mu\phi\partial^\mu\phi-m_\phi^2\phi^2)-\frac{g_{\phi\gamma}}{4}\int d^4x\,F_{\mu\nu}\tilde F^{\mu\nu}\phi,
\end{align}
where the first term is the usual QED kinetic term of the free electromagnetic field, the second term is a correction to the QED action due to vacuum 
polarization, namely, the Euler-Heisenberg term, the third one is the action of a free pseudoscalar field, $\phi$, with mass $m_\phi$, 
and the last one describes the interaction of the electromagnetic field with $\phi$. The quantities in Eq. \eqref{em-action} are defined as follows: $\alpha=1/137$ is the fine structure constant, $m_e$ is the electron mass, $F_{\mu\nu}$ is the electromagnetic field tensor, and 
$\tilde F_{\mu\nu}$ is its dual.
The coupled equations of motion\footnote{By equation of motion we mean classical equation of motion where only the first quantization is involved.} for the electromagnetic field tensor, $F_{\mu\nu}$, and for the field $\phi$ are, respectively, 
\begin{eqnarray}
\partial_\mu\left(F^{\mu\nu}-\frac{\alpha^2}{90m_e^4}[4F^2F^{\mu\nu}+7(F\tilde F)\tilde F^{\mu\nu}]\right) &= &-g_{\phi\gamma}\tilde F^{\mu\nu}\partial_\mu\phi,\label{gamma-eq-1}\\
(\Box+m_\phi^2)\phi=-\frac{1}{4}g_{\phi\gamma}F\tilde F\label{axion-eq-1},
\end{eqnarray}
where $F^2=F_{\mu\nu}F^{\mu\nu}$ and $F\tilde F=F_{\mu\nu}\tilde F^{\mu\nu}$. The electromagnetic field tensor is given in general by the sum of the incident photon field tensor $f_{\mu\nu}$ and of the external field tensor $F_{\mu\nu}^{(e)}$, $F_{\mu\nu}=f_{\mu\nu}+F_{\mu\nu}^{(e)}$. The
second term in the left-hand side of Eq. \eqref{gamma-eq-1} gives an effective mass to the photon in the presence of an external magnetic field:
 $m_\gamma^2=\omega^2-k^2\neq 0$, where $\omega$ is the photon energy and $k=n\omega$ is the photon momentum with $n$ being the total refraction index of the medium. In the limit where the coherence length of the external magnetic field, $\lambda_B$, is greater than the photon wavelength, $\lambda_p$; $\lambda_p\ll \lambda_B$, the coupled system of Eq. \eqref{gamma-eq-1} and Eq. \eqref{axion-eq-1}, 
 is well described by the WKB approximation and reduces to the following first order matrix equation \cite{Raffelt:1987im}:
\begin{equation}\label{matrix-eq}
\left[(\omega+i\partial_{\mathbf{x}})\mathbf I+
\begin{bmatrix}
  m_{+} & 0 & 0 \\
  0  &  m_{\times} & m_{\phi\gamma} \\
  0 & m_{\phi\gamma} & m_a\\
   \end{bmatrix}
\right] 
\begin{bmatrix}
  A_+\\
  A_\times\\
   \phi\\
 \end{bmatrix}
=0\,,
\end{equation}
where\footnote{$m_a$ should not be confused with the QCD axion mass, which here we denote as $M_a$ (see the Introduction). However, in this paper we do not consider the case of the QCD axion.} $m_+=\omega(n-1)_+$, $m_\times=\omega(n-1)_\times$, $m_{\phi\gamma}=g_{\phi\gamma}B_T/2$, $m_a=-m_\phi^2/2\omega$, $\mathbf I$ is the unit matrix, $B_T$ is the strength of the external magnetic field $\mathbf B_e$,
which is transverse to the direction, $\mathbf x$,  of the photon/axion propagation, and 
$A_{+, \times}$ are the photon polarization states with ${+, \times}$ being the polarization indexes (helicity) of the photon. 
The helicity state $+$ corresponds  to the polarization perpendicular to the external magnetic field and $\times$ describes the polarization parallel to the external field. The total index of refraction is given by the sum of three main components: the index of refraction due to electronic plasma $n_\textrm{pla}$, index of refraction due to vacuum polarization $n_\textrm{QED}$, and the one due to the Cotton-Mouton effect $n_\textrm{CM}$. The index of refraction due to electronic plasma is given by $n_\textrm{pl}-1=-\omega_\textrm{pla}^2/2\omega^2$ where 
{ in the limit of small photon frequencies, $\omega \ll m_e$}, $\omega_\textrm{pla}^2=4\pi n_e/m_e$ with $n_e$ being the number density of free 
electrons in the plasma. 
The index of refraction due to the QED effects, for $\omega\ll (2m_e/3)\times (B_c/B)$, is given by Ref. \cite{Tsai:1974fa}:
\begin{equation}\label{QED-index}
n_{\times, +} =1+\frac{\alpha}{4\pi}\left(\frac{B_T}{B_c}\right)^2\left[\left(\frac{14}{45}\right)_{\times}, \left(\frac{8}{45}\right)_{+}\right],
\end{equation}
where $B_c=m_e^2/e=4.41\times 10^{13}$ G. In principle one should also take into account the index of refraction of photons in the cosmological 
gas which is formed after the recombination epoch and mostly composed of neutral hydrogen atoms. 
The index of refraction of light in hydrogen gas is only measured in some restricted frequency range 
at zero degree Celsius and at normal pressure. Since the only epoch when the hydrogen gas was abundantly present in the Universe is 
confined between the re-ionization and recombination with the temperatures in the range\footnote{In this paper we do not consider 
the passage of the CMB photons through the interstellar medium which is mostly composed of hydrogen atoms.} from 
$T\sim$ 3000~K down to $20$ K, it may seem that using the measured value of the hydrogen index of refraction is not appropriate. However, as is demonstrated by
experiment and supported by the theory, the index of refraction very weakly depends upon the photon frequency for the frequencies
below the atomic level, so it is reasonable to use the experimental results in the cosmological conditions. Based on that we have checked that using the index of refraction of the hydrogen gas, as that presented in Ref. \cite{Mirizzi:2009nq}, does not affect our results at all.
Thus we do not take it into account in what follows. We also do not consider  the contribution of the Cotton-Mouton effect to the total index of refraction, since the Cotton-Mouton constant is very difficult to find in general\footnote{In general the Cotton-Mouton constant $C$ is measured for liquids and molecular gases such as H$_2$, O$_2$ etc. At the post recombination epoch the Universe is filled with atomic hydrogen H and helium He. Using the value of $C_\textrm{He}=2.19\times 10^{-16}$ T$^{-2}$ at $P=1$ atm for He as presented in Ref. \cite{Cadene2013} one can easily find that $n_\textrm{CM}\ll n_\textrm{pla}$ at post recombination epoch.} and in particular in cosmological situation considered in this paper. Indeed, the Cotton-Mouton constant depends on the absolute temperature of the gas, on the radiation wavelength, and on pressure, $P$. The latter, in the cosmological situation considered here is essentially zero, $P\simeq 0$, after recombination when the Universe is matter dominated. Therefore we expect the index of refraction due to the Cotton-Mouton effect to be very small and completely negligible with respect to the index of refraction of electronic plasma, $n_\textrm{CM}\ll n_\textrm{pla}$.

In order to take into account the interactions of photons and ALPs with the surrounding medium, we work with the 
density operator $\hat\rho$. It satisfies the well-known quantum kinetic equation
\begin{equation}
\frac{d\hat \rho}{dt}=-i[\hat M,\hat \rho]-\{\hat\Gamma, (\hat \rho-\hat \rho_\textrm{eq})\},
\label{eq-dens-matrix}
\end{equation}
where $\hat M$ is an operator (the Hamiltonian of the system) which takes into account the forward scattering (index of 
refraction) of photons and axions, $\hat\Gamma$ is an operator which takes into 
account non-forward scattering or decoherence effects and annihilation, 
and $\hat \rho_{eq}$ is the equilibrium density operator. It is diagonal and proportional to 
the unit matrix for equal mass fields. So it would be diagonal in any basis. However, for  
unequal masses the equilibrium density matrix is diagonal only in the mass (free 
Hamiltonian) eigenstate basis, while the damping coefficient matrix is diagonal in
the interaction (flavor) eigenstate basis.   
It is worth noting that the damping term (proportional to $\Gamma$) in Eq. \eqref{eq-dens-matrix} is an approximate and a more
precise description can be given in terms of the collision integral in the same way as it is done for oscillating neutrinos,
see e.g. the review~\cite{Dolgov:2002wy}, Eqs. (292)-(295). An exact equation describing the photon-ALP mixing 
will be presented in Ref.~\cite{Ejlli}.

From Eq. \eqref{matrix-eq} we can see that only the state $A_\times$ mixes with the pseudo-scalar field and for this reason it is better to work with a reduced density operator constructed from the two component field $\Psi^T=[A_\times, \phi]$, where $T$ means the transpose of a given matrix. In the basis spanned by the field $\Psi$, the matrix representation of the free Hamiltonian operator $\hat M$ and of the damping operator $\hat\Gamma$ are, respectively, given by
\begin{equation}
M=
\begin{bmatrix}
m_\times & m_{\phi\gamma}\\
m_{\phi\gamma} & m_a
\end{bmatrix},
\quad 
\Gamma=
\begin{bmatrix}
\Gamma_\gamma & 0\\
0 & \Gamma_\phi
\end{bmatrix},
\end{equation} 
where $M$ is a part of the Hamiltonian in the Schrodinger equation \eqref{matrix-eq} and $\Gamma_\gamma, \Gamma_\phi$ are, respectively, the coherence breaking interaction rates of photons and axions with the medium.

The evolution of the elements of the density operator $\hat\rho$ is determined by the equations:
\begin{eqnarray}\label{densitysys}
\rho_{\gamma}' &=&\frac{2m_{a\gamma}I + \Gamma_\gamma\, (\rho_{\gamma}-\rho_\textrm{eq}^\gamma)}{HT},\label{gamma-eq}\\
\rho_{\phi}' &=& \frac{-2m_{a\gamma}I+\Gamma_\phi(\rho_{\phi}-\rho_\textrm{eq}^\phi)}{HT} ,
\label{axion-eq}\\
R'&=& \frac{-(m_\times-m_a)I+(\Gamma_\gamma+\Gamma_\phi)R/2}{HT}\label{Real} ,\\
I'&=& \frac{(m_\times-m_a)R+(\Gamma_\gamma+\Gamma_\phi)I /2 + m_{\phi\gamma}(\rho_{\phi}-\rho_{\gamma})}{HT}\label{I},
\end{eqnarray}
where we split the off-diagonal terms of the density matrix into real and imaginary parts 
as $\rho_{\phi\gamma}=\rho_{\phi\gamma}^{*} =R+iI$,
$\rho_{eq}^\gamma=(e^{\omega/T}-1)^{-1}$, and $\rho_{eq}^\phi=(e^{\omega/T}-1)^{-1}$ are, respectively, the equilibrium distribution 
functions for photons and ALPs with zero chemical potential, $\mu=0$, and
the prime indicates the derivate with respect to the temperature, $T$. Indeed, we have expressed the total 
time derivative in Eq. \eqref{eq-dens-matrix}
as $d/dt=\partial_t-Hp\partial_p=Ha\partial_a=-HT\partial_T$, where $a$ is the cosmological scale factor, 
$p$ is the particle momentum, and $H$ is the Hubble parameter. 
Here we assumed that the temperature drops down as $T\sim 1/a$ and the derivative is taken
with respect to temperature, $T$, which is measured in Kelvin. The diagonal terms of the density operator are the usual occupation numbers, namely, $\rho_\gamma(\omega, T)\equiv n_\gamma(\omega, T)$ and $\rho_\phi(\omega, T)\equiv n_\phi(\omega, T)$. They should be not confused with the photon and ALP energy density $\rho_\gamma, \rho_\phi$ if not explicitly stated.
In terms of the Universe temperature $T$, the expressions for 
$m_{\phi\gamma}$, $m_\times$, and $m_a$ are given by
\begin{eqnarray}
m_{\phi\gamma}(T)& = & 1.12 \times 10^{-26}\left(\frac{B}{\textrm{nG}}\right)
\left(\frac{g_{\phi\gamma}}{10^{-11}\textrm{GeV}^{-1}}\right)\left(\frac{T} {T_0}\right)^2\quad \textrm{K}^{},
\label{m-a-gamma}\\
m_\times(T) &=& 7.13 \times 10^{-50} x_0 \left(\frac{B}{\textrm{nG}}\right)^2\,\left(\frac{T}{T_0}\right)^5 - 3.45 \times 10^{-14} x_0^{-1}
\left(\frac{n_e(T)}{\textrm{cm}^{-3}}\right)
\left(\frac{T_0}{T}\right)  \textrm{K},
\label{m-times}\\
m_a(T) &=& - 2.53 \times 10^{-3} x_0^{-1}\left(\frac{m_\phi}{10^{-5}\textrm{eV}}\right)^2
\left(\frac{T_0}{T}\right)\quad \textrm{K}^{},
\label{m-a}
\end{eqnarray}
where $x_0=\omega_0/T_0$ and $T_0=2.725$ K \cite{Komatsu:2010fb} and $\omega_0$ are, respectively, the present 
day CMB temperature and frequency/energy;  $x_0$ does not depend on the redshift and we often use $x\equiv x_0$. In Eq. \eqref{m-times} $m_\times$ is respectively the sum of QED and plasma contributions 
to the refraction index, namely, $m_\times=(m_\textrm{QED}+m_\textrm{pla})_\times$ for the photon polarization state $\times$.

\section{\label{sec:3}Coherence breaking terms in the cosmological plasma}

Equations of motions for the density operator, Eqs.  \eqref{gamma-eq}-\eqref{I}, include several terms which have not yet been 
calculated. These terms are the interaction rates of photons and ALPs with the medium, 
$\Gamma_\gamma$ and $\Gamma_\phi$, 
the plasma ionization fraction, $X_e$, which enters into expression for the free electron number density, $n_e$,  
and the factor $1/(HT)$. Once these terms are calculated, Eqs. \eqref{gamma-eq}-\eqref{I} are solved numerically and their solutions present the occupation number of photons and ALPs.

The term $1/(HT)$ can be calculated once we know the evolution  of the Hubble parameter, $H$, with temperature.
At the present epoch the Universe is dominated by the vacuum energy density and the matter energy density which are usually quantified in terms of their respective density parameters $\Omega_\Lambda$ and $\Omega_M$. Another contribution to the total energy density comes from relativistic particles such as CMB photons and the relic background of cosmological neutrinos\footnote{ Here we assume that all three neutrino species are nearly massless. In case neutrinos have mass above the eV scale their contribution to non relativistic matter at the present day can be easily taken into account. } which have a density parameter at 
present equal to $\Omega_R$. Including all these three contributions into the total energy density, we can write the Hubble parameter as follows: 
\begin{equation}\label{Hubble-par}
H^2=H_0^2\left[\Omega_\Lambda+\Omega_M\left(\frac{T}{T_0}\right)^3+\Omega_R\left(\frac{T}{T_0}\right)^4\right] ,
\end{equation}
where we used the Friedmann equation $H^2=(8\pi G/3)\rho$, where $G$ is the Newtonian constant, $\rho=\sum_i\rho_i$ is the total energy density\footnote{In this section $\rho$ represents the cosmological energy density
and should not be confused with the density operator in the previous section.} of the Universe with $i=\{\Lambda, M, R\}$, $\Omega_i=\rho_i/\rho_c$, and $\rho_c=3H_0^2/8\pi G=1.878\times 10^{-29}$ g/cm$^3$ is the critical energy density at the present epoch. The present day value of the Hubble parameter is $H_0=100 h_0$ km/s/Mpc where $h_0=0.6711$ according to the Planck Collaboration \cite{Ade:2013zuv}. Also, according to the Planck results,
the values of the density parameters of vacuum energy and  
 nonrelativistic matter are, respectively, $\Omega_\Lambda=0.68$ 
 and $\Omega_Mh_0^2=0.12$. The density parameter of the CMB photons can be calculated using the Stefan-Boltzmann law for  the equilibrium 
 radiation at  temperature $T=2.725$ K
\begin{equation}
\rho_\gamma(T_0)=\sigma T_0^4=4.64\times 10^{-34} \textrm{g cm}^{-3},
\end{equation}
where $\sigma=7.56\times 10^{-15}$ erg cm$^{-3}$ K$^{-4}$ is the Stefan-Boltzmann constant. Therefore, the density parameter of the CBM photons is given by
\begin{equation}
\Omega_\gamma(T_0)=\frac{\rho_\gamma(T_0)}{\rho_c}=2.47\times 10^{-5}h_0^{-2}.
\end{equation}
The neutrino energy density is suppressed with respect to $\rho_\gamma$ because of the larger temperature of photons due to their heating by $e^+e^-$-annihilation after neutrino decoupling at $T\simeq 10^{10}$ K. Correspondingly the ratio of the neutrino 
temperature, $T_\nu$, to the photon temperature, $T_\gamma$, becomes $T_\nu/T_\gamma=(4/11)^{1/3}$. Since photons and neutrinos are relativistic particles, their energy densities scale with temperature as $\rho_R\sim T^4$. Taking into account that neutrinos are fermions and that there are three neutrino species, we find that the total contribution of neutrinos to the energy density of the relativistic matter is
\begin{equation}
\rho_\nu=3\times \frac{7}{8}\left(\frac{4}{11}\right)^{4/3}\rho_\gamma,
\end{equation}
where the factor $7/8$ comes from the Fermi-Dirac statistics. Thus the total density parameter of relativistic species at the present 
time is
\begin{equation}
\Omega_R=\left[1+3\times \frac{7}{8}\left(\frac{4}{11}\right)^{4/3}\right]\Omega_\gamma(T_0)=4.15\times 10^{-5}h_0^{-2}.
\end{equation} 
Inserting all the necessary quantities into Eq. \eqref{Hubble-par}, we obtain the following expression for:
\begin{equation}\label{Inverse-T}
\frac{1}{HT}=2.18\times 10^{28}\left[0.68\left(\frac{T}{T_0}\right)^2+0.26\left(\frac{T}{T_0}\right)^5+9.21\times 10^{-5}\left(\frac{T}{T_0}\right)^6\right]^{-1/2}\quad \textrm{K}^{-2}.
\end{equation}
In Eq. \eqref{Inverse-T}, only the contribution of neutrinos and photons to the relativistic energy density, is included. Since ALPs are also supposed to be relativistic, $m_\phi<T_0$, their energy density is to be taken into account as well. This would be so, if primordial ALPs existed in the plasma. We, however, assume that the only source of ALPs is their production by photon transformation in the magnetic field. Since energy is conserved in this process, there is no change in the total energy density of the relativistic species. 

Another important term which we need in order 
to solve Eqs. \eqref{gamma-eq}-\eqref{I} is the ionization fraction $X_e(T)$ which gives the fraction of free electrons at the post BBN epoch. At temperatures below $T\leq 4226$ K the ionization fraction in a very good approximation is
determined by the following differential equation (see Ref.  \cite{Weinberg:2008} for further details and references):
\footnote{More precisely, the differential equation \eqref{ionization}  for the ionization fraction is based on the so-called hydrogen three level approximation.} 
\begin{equation}\label{ionization}
\frac{dX_e}{dT}=\frac{\alpha n}{HT}\left(1+\frac{\beta}{\Gamma_{2s}+8\pi H/\lambda_{\alpha}^3n_B(1-X_e)}\right)^{-1}\left(\frac{SX_e^2+X_e-1}{S}\right),
\end{equation}
where $\Gamma_{2s}=8.22458$ s$^{-1}$ is the two-photon decay rate of the $2s$ hydrogen state, 
$\lambda_{\alpha}=1215.682\times 10^{-8}$ cm is the wavelength of the Lyman $\alpha$ 
photons,  $\alpha(T)$ is the case B (see Ref. \cite{Weinberg:2008})  recombination coefficient, 
and $S(T)$ is the coefficient in the Saha equation, $X_e(1+SX_e)=1$. 
The function $S(T)$ is given by 
\begin{equation}
S(T)=1.747\times 10^{-22}e^{157894/T}\left(\frac{T}{1 \textrm{K}}\right)^{3/2}(\Omega_Bh_0^2),
\end{equation}
where $\Omega_Bh_0^2=0.022$ is the baryon density parameter. 
The functions $\alpha(T)$ and $\beta(T)$ are given, respectively, by Refs. \cite{Hummer1994}:
\begin{eqnarray}
\alpha(T) &=& \frac{1.4377\times 10^{-10}\left(\frac{T}{1 \textrm{K}}\right)^{-0.6166}}{1+5.085\times 10^{-3}\left(\frac{T}{1 \textrm{K}}\right)^{0.53}}\quad \textrm{cm}^3\, \textrm{s}^{-1},\\
\beta(T) &=& 2.4147\times^{15}\left(\frac{T}{1 \textrm{K}}\right)^{3/2}e^{-39474/T}\alpha(T)\quad \textrm{cm}^{-3}.
\end{eqnarray}
With these parameters, Eq. \eqref{ionization} can be solved numerically by imposing the initial condition $X(4226)=0.98$ for $\Omega_Mh_0^2=0.12$, see Ref. \cite{Weinberg:2008} for more details. The solution of Eq. \eqref{ionization} gives the ionization fraction of 
the hydrogen atoms for the temperature $T\leq 4226$ K. For temperature higher than $T> 4226$ K  the hydrogen atoms are completely ionized and the hydrogen ionization fraction is unity.

\begin{figure}[htbp]
\begin{center}
\includegraphics[scale=0.8]{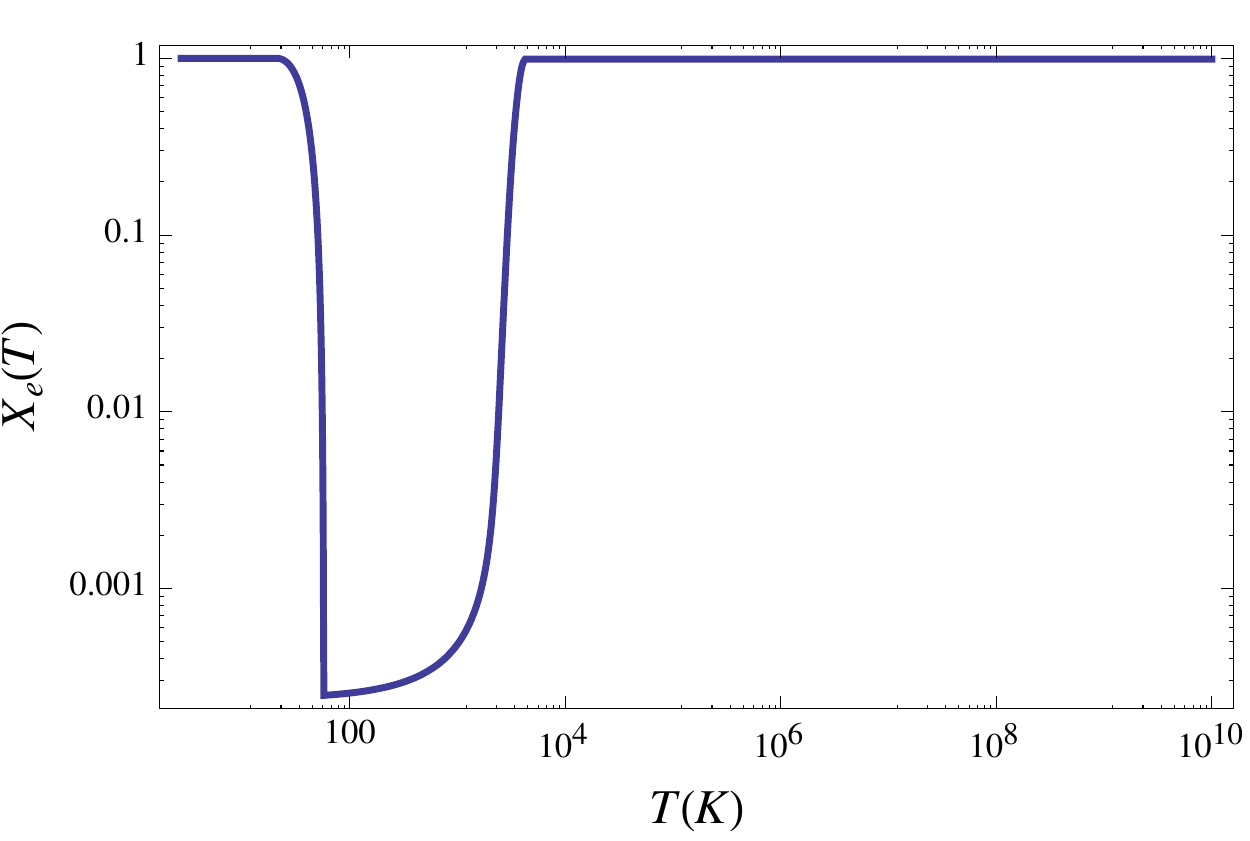}
\caption{Plot of the hydrogen ionization fraction $X_e(T)$ as a function of temperature, $T$, for $\Omega_Bh_0^2=0.022$ and $\Omega_Mh_0^2=0.12.$}
\label{plot-ionization}
\end{center}
\end{figure}

The solution of Eq. \eqref{ionization}  for $X_e(T)$ is valid until the reionization
time, when complete ionization is adiabatically restored, i.e.
$X_e$ reaches unity again. In Fig. \ref{plot-ionization} the hydrogen ionization 
fraction as a function of temperature is shown. In the temperature interval 
$57.22$ K$\leq T\leq 4226$ K the curve is obtained by the solution of 
Eq.~\eqref{ionization}, where the lower limit $T=57.22$ K corresponds 
to the \emph{start} of reionization  at $z_\textrm{ion}\sim 20$ with $z$ 
being the redshift with respect to the present time.
The complete reionization is reached at $z_\textrm{ion}\sim 7$ \cite{Dunkley:2008ie}. 
The evolution of  $X_e(T)$ in the temperature interval $21.8$ K $\leq T\leq 57.22$ K has been obtained by a smooth interpolation of the curve $X_e(T)$ in the interval 
$57.22$ K$\leq T\leq 4226$ K with $X_e=1$ of the interval $2.725$ K$\leq T\leq 21.8$ K. 

Both photons and ALPs scatter in the surrounding medium which is mostly composed of
free protons, electrons, and atomic nuclei. For the CMB photons an important 
coherence breaking term comes from the scattering on free electrons. 
Indeed, the CMB photons are observed at present by COBE \cite{Fixsen:1996nj} in 
the energy range $1.2\, T_0\leq\omega_0\leq 11.3\,T_0$ and at the post BBN epoch 
their energy is smaller than the electron mass, $\omega<m_e$. In this case the interaction rate of 
photons, $\Gamma_\gamma$, is given by
\begin{equation}\label{Com}
\Gamma_\gamma(T)=\sigma_{T}n_e(T)
\end{equation}
where $\sigma_{T}$ is the Thomson total cross section, $\sigma_T=6.65\cdot 10^{-25}$ cm$^2$. 
The free electron density, $n_e$, can be written in general as 
\begin{equation}\label{free-ele}
n_e(T)=\left(Y_\textrm{H}X_e+\sum_jZ_jX_e^jY_j \right) n_B(T),
\end{equation}
where $X_e^j$ is the ionization fraction of the light elements such as He$^4$, He$^3$ etc, $n_B(T)$ is the total baryon number density which is given by $n_B(T)=n_B(T_0)(T/T_0)^3$, where according to the Planck Collaboration~\cite{Ade:2013zuv} $n_B(T_0)\simeq 2.47\times 10^{-7}$ cm$^{-3}$, $Z_j$, is the atomic number of the $j$-th specie and $Y_j$ is the fractional abundance by weight. As we already have seen, hydrogen is completely ionized at $T>4226$ K and its abundance 
by weight is $Y_\textrm{H}\simeq 0.76$. On the other hand the ionization of helium happens to be at $T\geq 2\times 10^4$ K and its abundance 
by weight is approximately $Y_\textrm{He}\simeq 0.24$. So we have $n_e(T)\simeq 0.76 X_e(T)n_B(T)$ for $T<2\times 10^4$ K and $n_e(T)\simeq 0.88 X_e(T)\,n_B(T)$ for $T\geq 2\times 10^4$ K. Equation \eqref{free-ele} is valid at 
the temperatures below the temperature when  $e^+e^-$-annihilation froze, 
$T_{e^\pm}\simeq m_e/10\simeq 5.7\times 10^8$ K. 
Above $T_{e^\pm}$ but below the electron mass,
$T<m_e$, the free electron density in thermal equilibrium is approximately  given by \cite{Kolb:1990vq}:
\begin{equation}
n_{e^-}(T)\simeq 2\left(\frac{m_eT}{2\pi}\right)^{3/2}e^{-m_e/T}=2.09\times 10^{16} e^{-m_e/T}\left(\frac{T}{T_0}\right)^{3/2}\quad \textrm{cm}^{-3}.
\label{eq-dens}
\end{equation}
Chemical potential in Eq. \eqref{eq-dens} is neglected because the asymmetry between electrons and  positrons is 
known to be very small. According to this equation there is
a huge increase of the free electron and positron density at $T>T_{e^\pm}$ which has an important impact on the 
transformation of photons into ALPs.

Apart from the dominant Compton scattering, there are three other coherence breaking terms, namely, the double 
Compton (dC) absorption ($\gamma_0+e\leftarrow \gamma_1+e+\gamma_2$), the thermal Bremsstrahlung absorption($\gamma+e+Ze\rightarrow e+Ze$), and the Primakoff effect ($\gamma+Ze\rightarrow \phi+Ze$). These processes do not 
conserve the CMB photon number and are essential for an elimination of the photon chemical potential, while the Compton scattering is not effective for that. The rate for the dC process  is given in Ref. \cite{Lightman1981}:
\begin{equation}\label{dC}
\Gamma_{dC}=\left(\frac{4\alpha}{3\pi}\right)\left(\frac{T_e}{m_e}\right)^{2}x_e^{-3}\left(e^{x_e}-1\right)\,g_{dC}(x_e)\Gamma_\gamma(T),
\end{equation} 
where $x_e=\omega/T_e$ with $T_e$ being the electron temperature in thermal equilibrium, $g_{dC}$ is the gaunt factor 
 depending on the ratio of temperature to energy, and $\Gamma_\gamma (T)$ is given by Eq. (\ref{Com}). 
 Close to thermal equilibrium 
 we can take $T_e\simeq T$ and $x_e\equiv x$. Gaunt factors are in general difficult to calculate and one often uses for them approximate fitting results. In the case of the double Compton process, 
 $g_{dC}$ is approximately given by Ref. \cite{Chubla2005}
\begin{equation}
g_{dC}(x_e)\simeq \frac{I}{1+14.16\,(T/m_e)}\,G_{dC}(x_e),
\end{equation}
where $I=\int dx\, x^4 n_{eq}(1+n_{eq})=4\pi^4/15$ and for the photon spectrum close to equilibrium:
\begin{equation}
G_{dC}(x_e)\simeq G_{dC}^{eq}(x_e)\simeq e^{-2x_e}\left(1+\frac{3}{2}x_e +\frac{29}{24}x_e^2+\frac{11}{16}x_e^3+\frac{5}{12}x_e^4\right).
\end{equation}

Another important coherence breaking term is the free-free absorption or thermal bremsstrahlung  which reduces the photon number. The absorption rate due to bremsstrahlung  in non-relativistic case $\omega<m_e$ is given by Refs. \cite{Lightman1981, Zeldovich:1969ff}\begin{equation}\label{brem}
\Gamma_{br}=\frac{\alpha}{\sqrt{24\pi^3}}\left(\frac{m_e}{T}\right)^{1/2} \frac{1}{\omega^3}(1-e^{-x_e}\,)\sum_jZ_i^2n_jg_{br}(x_e, Z_j)\Gamma_\gamma(T),
\end{equation}
where $g_{br}(x_e)$ is the gaunt factor for thermal bremsstrahlung, $n_j$ is the number density of the $j$-th ion species with atomic number $Z_j$. The plasma at $z>1090$ is mainly composed of hydrogen and helium nuclei and 
we can write $\sum_j Z_j^2n_jg_{br}\simeq n_Bg_{br}$, where the helium and hydrogen Gaunt factors are approximately 
equal, $g_{br}$(H)$\simeq g_{br}$(He). In principle one should take into account the full expressions for hydrogen and helium Gaunt factors, however, as we have explicitly checked, the numerical results presented in the next sections weakly 
depend on their full expression. Therefore from now on we consider hydrogen and helium Gaunt factors as equal. 
In this case the analytic expression for $g_{br}(x_e)$ in the region $x_e\geq 1$ is 
given by Refs.~\cite{Lightman1981, Zeldovich:1969ff}:
\begin{equation}
g_{br}(x_e) =\ln(2.2)\,x_e^{-1/2}
\end{equation}

In the Primakoff effect, photons can transform into ALPs in the Coulomb field of nuclei and electrons. 
In this case the interaction rate has two different expressions in the limits of $T<m_e$ and $T> m_e$. 
Since we are interested in the post BBN epoch, $T<1$ MeV, the absorption rate in the 
non relativistic case, $T< m_e$, is given by Ref. \cite{Raffelt:1985nk}:
\begin{equation}\label{Gamma-a}
\Gamma_{\gamma\rightarrow a}=\frac{g_{\phi\gamma}^2Tk_c^2}{32\pi}\left[\left(1+\frac{k_c^2}{4\omega^2}\right)\log\left(1+\frac{4\omega^2}{k_c^2}\right)-1\right],
\end{equation}
where $k_c$ is a cutoff momentum which takes into account the Debye
screening effects in the plasma:
\begin{equation}
k_c^2=\frac{4\pi\alpha}{T}\left( n_e+\sum_jZ_j^2n_j \right).
\end{equation}
However, we have checked that the absorption rate due to the Primakoff effect is completely negligible in comparison with the double Compton and bremsstrahlung absorption rates in the ALP parameter space considered in this paper. Consequently, in what follows we neglect the Primakoff effect.

In the case of ALPs the most important coherence breaking effect comes from the decay of ALPs into two 
photons $\phi\rightarrow \gamma+\gamma$ with the rate 
\begin{equation}\label{axion-decay}
\Gamma_{\phi\rightarrow 2\gamma}=\frac{g_{\phi\gamma}^2m_{\phi}^3}{64\pi}=
2.51\times 10^{-53}\left(\frac{g_{\phi\gamma}}{10^{-11}\textrm{GeV}^{-11}}\right)^2
\left(\frac{m_{\phi}}{10^{-5}\textrm{eV}}\right)^3\quad \textrm{cm}^{-1}.
\end{equation}
Since in this paper we consider only relativistic ALPs with $m_\phi\ll T_0$ and the expected coupling constant is $g_{\phi\gamma}<10^{-11}$ GeV$^{-1}$ (see the next sections), the ALP lifetime $\tau=1/\Gamma_{\phi\rightarrow 2\gamma}$ is in general much longer than the Universe's age and therefore the decay of ALPs into photons can be neglected. With the interaction rates presented above our density matrix formalism is very general and applicable for $\omega<m_e$.

\section{\label{sec:4} ALP production at the post BBN epoch. Analytic solution.}

Here we present an approximate solution of Eqs. (\ref{gamma-eq}) -(\ref{I}) using
the steady state approximation. This approximation is valid when the oscillation length of the photon-ALP system is 
shorter than the Hubble 
distance, $l_{osc}\ll H^{-1}$, which is well satisfied at the post BBN epoch. 
In this case the terms in the right hand sides of Eqs. \eqref{Real} and \eqref{I} multiplied by the large factor $1/(HT)$
must be  approximately zero and expressing $R$ and $I $ through the diagonal components of the density matrix we obtain
\be
R &=& \frac{ 4 \Delta m\, m_{\phi\gamma}}{4\Delta m^2+\Gamma^2}(\rho_{\gamma}-\rho_{\phi})  ,
\label{R-sol}\\
I & = & \frac{ 2 m_{\phi \gamma} \Gamma} {  4 (\Delta m)^2+\Gamma^2}(\rho_{\gamma} - \rho_{\phi}), 
\label{I-sol}
\ee
where $\Delta m = m_\times - m_a$ and 
$\Gamma = \Gamma_\gamma + \Gamma_\phi$.
Using Eqs. \eqref{gamma-eq} and \eqref{axion-eq} we find:
\be
\rho'_{\gamma}& = & \frac{1}{HT}\left[\Gamma_\gamma(\rho_{\gamma} - 
\rho_{\gamma}^{eq}) + \frac{ 4 \Gamma m_{\phi\gamma}^2}{ 4\Delta m^2 + \Gamma^2}(\rho_{\gamma}
-\rho_{\phi}) \right] 
\label{rho-gamma-prime},\\
\rho'_{\phi}& = & \frac{1}{HT}\left[  \Gamma_\phi (\rho_{\phi} - 
\rho_{\phi}^{eq}) - \frac{ 4\Gamma m_{\phi\gamma}^2}{ 4\Delta m^2 + \Gamma^2} (\rho_{\gamma}
-\rho_{\phi}) \right] 
\label{rho-gamma-prime1} .
\ee
The effective reaction rate of the ALPs, $\Gamma_\phi$ for the ALP mass range considered here is tiny, so
we take  $\Gamma_\phi \simeq 0$. If the photon occupation number is close to the equilibrium one, 
$\rho_{\gamma} \approx \rho_{\gamma}^{eq} $ and $\rho_\gamma\gg \rho_\phi$, the equation for
$\rho_{\phi}$ simplifies to
\be
P_{\phi}'\equiv\frac{\rho'_{\phi}}{\rho_\gamma^{eq}} \simeq -\frac{ 4 m_{\phi\gamma}^2 \Gamma}
{ HT (4\Delta m^2 + \Gamma^2)}.
\label{rho-aa-simple}
\ee

We solve Eq. \eqref{rho-aa-simple} when the transition is dominated by the 
resonance. The latter occurs when $\Delta m = 0$. The ALP production probability, $P_\phi$, at the resonance can be calculated as follows. We expand 
$\Delta m$ near resonance temperature as $\Delta m = \kappa (T-\bar T)$, where
 $\kappa = d(\Delta m)/dT$ and $\bar T$ is the resonance temperature. After that we integrate around the 
resonance temperature and obtain 
\begin{equation}\label{axion-res}
P_{\phi} =-\left.\frac{2\pi m_{\phi\gamma}^2 }{\kappa HT}\right|_{T=\bar T},
\end{equation}
where in Eq. \eqref{axion-res} each quantity must be evaluated at the resonance temperature, $\bar T$. In order to proceed on the calculation of $P_\phi$ at the resonance temperature it is convenient to express, $\Delta m$ as follows
\begin{equation}
\Delta m(T)=m_\textrm{QED}(T)-m_\textrm{pla}(T)-m_a(T)
\end{equation}
Keeping in mind that $\kappa(T)=(5m_\textrm{QED}(T)-2m_\textrm{pla}(T)+m_a(T))/T$ and that $\Delta m(\bar T)=0$, we get the following expression for the function $\kappa (T)$ at the resonance
\begin{equation}\label{k-resonance}
\kappa(T)=\frac{3}{T}\left.\left(2m_\textrm{QED}(T)-m_\textrm{pla}(T)\right)\right|_{T=\bar T},
\end{equation}
where we assumed $X_e'(T)=0$. Inserting Eq. \eqref{k-resonance} into Eq. \eqref{axion-res} we get 
\begin{equation}\label{rho-res}
P_\phi(T)=\left. -\left(\frac{2\pi}{3H}\right)\frac{m_{\phi\gamma}^2}{m_\textrm{a}+m_\textrm{QED}}\right|_{T=\bar T}.
\end{equation}

\begin{figure}[h!]
\begin{center}
\includegraphics[scale=0.7]{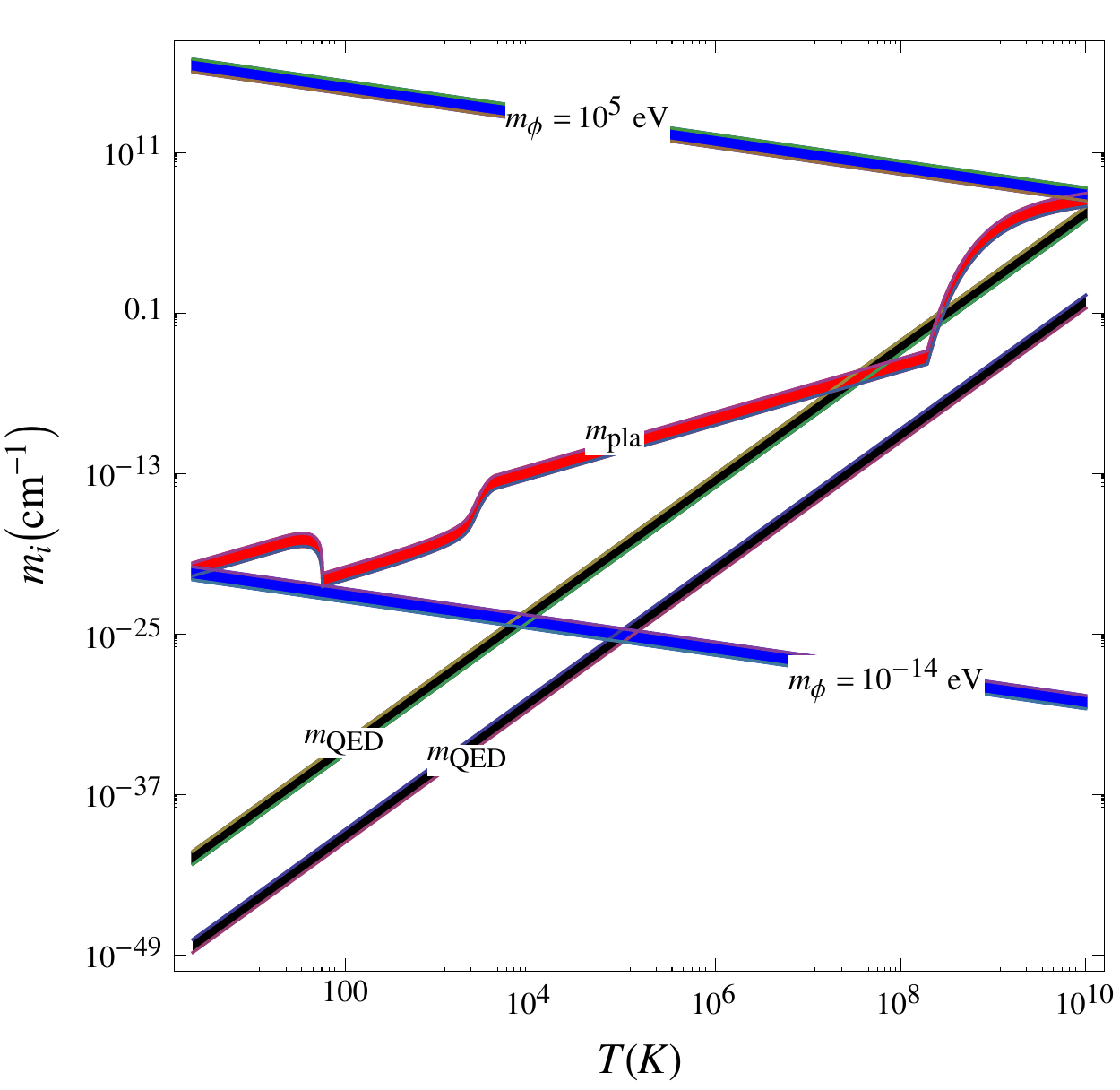}
\caption{Log-Log plot of  $m_i=\{m_\textrm{pla}, m_\textrm{QED}, m_a\}$ as a function of  temperature, $T$, in Kelvin 
in the CMB energy range $1.2\leq x\leq 11.3$ (colored regions). The QED terms, $m_\textrm{QED}$ (in green upper region and in black lower region) for $B=2\times 10^3$ nG and $B=1$ nG and plasma term $m_\textrm{pla}$ (in red) are respectively shown. The term $m_a$ (in blue or in black) for fixed values of $m_\phi=10^{5}$ eV (top region) and $m_\phi=10^{-14}$ eV (bottom region) is shown. {{The term $m_a$ and plasma term $m_\textrm{pla}$ do not depend on  $B$.}}}
\label{Fig2a}
\end{center}
\end{figure}

Equation \eqref{rho-res} allows us to constrain the ALP parameter space in the resonant case, e.g. at the present time. Clearly, this can be done at other cosmological epochs as well.  First let us  
note that  for $m_\phi>10^{-15}$ eV we have  $m_a\gg m_\textrm{QED}$ when $T<10^4$ K.  Indeed, 
this can be seen from Fig. \ref{Fig2a} where  $m_i(T)$ as a function of temperature is shown. 
The term, $m_a$ for $m_\phi=10^{-14}$ eV (bottom region in blue) and $m_\phi=10^{5}$ eV (top region in blue) are 
presented  for $1.2\leq x\leq 11.3$. Thus, as far as $m_\phi> 10^{-14}$ eV, the  QED effects are completely negligible 
at the post recombination epoch.

Based on this fact we can neglect the QED term in Eq. \eqref{rho-res} in comparison to $m_a$. In this case the resonance probability reads
  \begin{equation}\label{rho-res1}
P_\phi(T)=\left. 4.57\times 10^{-49}\,x\,H^{-1}\left(\frac{10^{-5} \textrm{eV}}{m_\phi}\right)^{2}\left(\frac{B}{\textrm{nG}}\right)^2\,\left(\frac{g_{\phi\gamma}}{10^{-11}\textrm{GeV}^{-1}}\right)^2\left(\frac{T}{T_0}\right)^5\right|_{T=\bar T}\quad \textrm{cm}^{-1}.
\end{equation} 
At this point we need to calculate the resonance temperature as a function of the ALP parameters. In the case of low temperatures $T<3000$ K we can neglect the QED term in $\Delta m$ and solve the equation for $\Delta m(T)=0$ to find the resonance temperature:
\begin{equation}\label{res-T}
\frac{\bar T}{T_0}=7.31\times 10^5 \left(\frac{\bar m_\phi}{10^{-5} \textrm{eV}}\right)^{2/3} X_e^{-1/3},
\end{equation}
{{where $\bar m_{\phi}$ is the the value of the ALP mass which solves the equation $m_a(\bar T, \bar m_\phi)=0$ and we call it the resonance mass}}. Imposing the condition that resonance happens at the
present epoch, $\bar T=T_0$ we obtain the following value for $\bar m_\phi$:
\begin{equation}\label{res-mass}
\bar m_\phi=1.6\times 10^{-14}X_e^{1/2} \textrm{eV}=1.6\times 10^{-14} \textrm{eV},
\end{equation}
where we took $X_e=1$ at the present epoch. Let us stress that we assumed that the ionization fraction $X_e$ is constant around the present epoch, $z<10$. However, if $X_e$ is not constant in the temperature interval under consideration, then Eq. \eqref{res-T} is an equation to be solved for $T$ where in this case $X_e(T)$ is a rather complicated function of the temperature. Since, we imposed the condition that the
resonance transition happens at the present epoch, when $\bar T=T_0$, we can take for 
$H(\bar T)^{-1}=H_0^{-1}=1.38\times 10^{28}$ cm, $\bar m_\phi$ given by Eq. \eqref{res-mass}, and then 
substitute the parameter values into Eq. \eqref{rho-res}. Demanding that $P_\phi(x)\lesssim F$ (where $F$ is a generic function which may depend on $x$ and either represent an upper bound from $\mu$-distortion or temperature anisotropy, see next section), we get the following upper bound  for $g_{\phi\gamma}B$:
\begin{equation}
g_{\phi\gamma}B\lesssim 2\times 10^{-10}\left(\frac{F}{x}\right)^{1/2}\quad \textrm{nG}\times \textrm{GeV}^{-1},
\label{prob-alp}
\end{equation}
for $\bar m_\phi=1.6\times 10^{-14}$ eV. Taking the upper value for CMB energy, $x=11.3$, and using, for example 
$F\simeq \delta T/T=10^{-5}$ for the temperature anisotropy, we get a tighter bound: 
$g_{\phi\gamma}B\leq 1.8\times 10^{-13}\quad \textrm{nG}\times \textrm{GeV}^{-1}$.

\section{\label{sec:5}CMB distortions and ALP production. Numerical results}

In the previous section we estimated the production probability in the resonant case and derived constrains on the coupling constant,  $g_{\phi\gamma}$.  
However, analogous estimates in nonresonance case is less precise, in particular because we assumed that $\rho_\gamma\simeq \rho_{eq}$, 
while $\rho_\phi\ll \rho_\gamma$. To avoid these limitations we relax, the assumptions
of the previous section and present here numerical solutions of kinetic equations in both \emph{resonant} and \emph{nonresonant} cases  in the steady state approximation. Probably the latter is the main source
of error but still one may expect that it gives reasonably accurate results.

In this section we present the limits on the ALP parameter space from 
the CMB spectral distortion and temperature anisotropy. It is instructive to describe, in particular, the quantitative behavior of spectral distortion at different redshifts
in order to make our results more clear. According to COBE measurements CMB has the 
Planck spectrum to a very high degree of accuracy in the observed energy interval $1.2\leq x\leq11.3$. 
COBE tightly constrained CMB spectral distortions and obtained stringent upper limits on the chemical potential, 
$|\mu|\lesssim 9\times 10^{-5}$, and the Compton parameter,  
$|y|\lesssim 1.5\times 10^{-5}$ at $95\%$ CL~\cite{Fixsen:1996nj}

The distortion of equilibrium CMB spectrum has been usually studied in the models with an energy influx  from some 
hypothetical source into the CMB thermal bath, such as electromagnetic  decays of hypothetical (dark matter) particles, etc.
The size of the effect depends upon the competition between the power of the source which creates nonequilibrium photons 
and efficiency of the reactions between photons and cosmic electrons, which push the system to
thermal equilibrium.  The effects of such equilibrium restoration reactions depend upon the  cosmological redshift when the
reactions took place. The essential reactions studied in the literature are the elastic Compton scattering, the inelastic double Compton scattering with emission or absorption of photons, and free-free emission or bremsstrahlung.

For high redshifts, $z\geq 2\times 10^6$~\cite{Chubla2005}, a powerful energy injection is allowed, with the contribution
of nonequilibrium energy of the same order of magnitude as the energy density of CMB, without causing a 
noticeable spectral distortion. 
For an early discussion of CMB spectral distortion, see Ref. \cite{Zeldovich:1969ff} and for further developments 
and possibile distortion mechanisms, see Refs. \cite{CMB-spectral}.

Comparing Eqs. \eqref{Com}, \eqref{dC}, and \eqref{brem} we see 
that at high redshifts Compton scattering and double Compton scattering/absorption
dominates over bremsstrahlung but at low energies  
the last two processes became more efficient. 
As far as  $z$ remained larger than $2\times 10^6$~\cite{Chubla2005}, double Compton scattering would be responsible for absorbing or creating photons at low energies, while Compton scattering redistributed them in energy. The overall effect would be that of the equilibrium black body distribution even in the presence of a large energy injection into the plasma comparable with the CMB energy density.

If energy is injected at intermediate redshifts, $2.88\times 10^5\lesssim z\lesssim 2\times 10^6$, the double Compton scattering and bremsstrahlung would be efficient only at low energies 
and could not create enough photons to produce the black body spectrum except for low energies. However, the Compton scattering was still efficient in maintaining kinetic equilibrium and any kind of the energy release 
would establish a Bose-Einstein spectrum with chemical potential $\mu > 0$. 
In the case of the energy injections at 
$z\lesssim 1.5 \times 10^4$, the Compton scattering cannot establish kinetic equilibrium anymore. So spectrum 
distortion at high energies  would would remain intact. An energy injection at low energies would be thermalized due to an enhanced role of the bremsstrahlung. Any energy injection for $z\lesssim 1.5 \times 10^4$ would distort the CMB spectrum, creating the so-called $y$-distortion.

In the case of the transition of the CMB photons into ALPs, energy is not injected into CMB but, vice versa, the energy 
is taken out. The probability of the transition depends upon the photon frequency, so we would expect a frequency 
dependent deficit of photons. However, as it is argued above, the spectrum distortion at $z> 2\times 10^{6}$ would be
smoothed down by electron-photon reactions in the plasma and the equilibrium black body spectrum would be
restored. For $ z\lesssim 2\times 10^{6}$ the spectral distortion induced by CMB transformation into ALPs is affected by two
phenomena. First, there is the mentioned above
primary spectral distortion due to the leakage of  energy from CMB into ALPs.
This effect results in a decrease of occupation number $n_\gamma(\omega)$ varying with $\omega$. 
Consequently the CMB total energy and number density would decrease and the photon temperature would 
tend to become slightly lower than the electron temperature, namely, $T\lesssim T_e$. 
The secondary spectral distortion arises when Compton scattering (when efficient) up scatters 
photons to higher energies because $T\lesssim T_e$ (after the transition) and an additional distortion of the spectrum would 
appear.

In order to use the COBE limits on chemical potential $\mu$, we need to relate photon occupation 
number $n_\gamma(x)$ (in this section we use the notations $n_\gamma (x)$, $n_\phi(x), n_{eq}(x)$ instead of 
$\rho_\gamma, \rho_\phi$ and $\rho_{eq}$) with $\mu$. In general the occupation number for a Bose-Einstein distribution is given by
\begin{equation}\label{occupation-n}
n_\gamma(x)=\frac{1}{e^{x+\mu}-1}.
\end{equation} 
If $\mu\ll 1$, it is sufficient to expand Eq. \eqref{occupation-n} only up to the first order in $\mu$:
\begin{equation}
n_\gamma\simeq n_{eq}+\mu\partial_xn_{eq}=n_{eq}-n_{eq}(1+n_{eq})\mu.
\end{equation}
The relative deviation from  equilibrium of the CMB photons, therefore can be written as:
\begin{equation}
\frac{n_{eq}-n_\gamma}{n_{eq}}\simeq(1+n_{eq})\mu.
\label{delta-n}
\end{equation}
The equilibrium value of the chemical potential of photons must be positive to ensure the positive definiteness of $n$. However, in the 
considered case equilibrium is not reached and so
$\mu$ is not necessarily constant but may depend upon the photon frequency, satisfying the condition $x +\mu(x) > 0$ to ensure $n>0$. If, however, in all frequency ranges $n_\gamma< n_{eq}$, then $\mu$ must be
positive for all frequencies \footnote{The ''chemical potential'' $\mu$ considered here is a dimensionless quantity and is related to thermodynamical chemical potential $\mu_{\textrm{ther}}$ as $\mu=-\mu_{\textrm{ther}}/T$. }.

At low redshifts considered here, the photon loss into ALPs is not restored by double Compton scattering and bremsstrahlung, so the deviation of the photon distribution from the equilibrium can be expressed
through the probability of the photon to ALP transformation as: $P_{\phi}=n_\phi/n_{eq}\simeq (n_{eq}-n_\gamma)/n_{eq}$.
Using Eq.~\eqref{delta-n} we can express $P_\phi$ through the chemical potential $\mu$
as:
\begin{equation}\label{mu}
P_{\phi}(x)\simeq \mu(x)f(x),
\end{equation}
where $ f(x)={e^x}/{(e^x-1)}$ and  we omit an indication to dependence of $P_{\phi}$ on $B$ and $m_\phi$.

Late time constraints on the ALP parameter space can be obtained by demanding that 
the transition probability of the CMB photons into ALPs 
does not exceed the observed temperature anisotropy, $P_\phi\lesssim \delta T/T$. Indeed an
existence of the large scale magnetic field at the post recombination epoch would create a deficit of the CMB photons in the 
direction perpendicular to $\bf B_e$ relative 
to photons propagating along the field direction. Therefore for $z<1090$ we demand that
\begin{equation}\label{temp-anis}
P_\phi(x)\lesssim \delta T/T,
\end{equation}
where $P_\phi$ depends only on the transverse part of the magnetic field strength. 
Limits from $y$-distortion will not be considered here, see the next section for discussion.

As is already mentioned, the chemical potential $\mu$ weakly depends on the photon energy for $x>1$, 
which is the region explored by COBE and the planned space missions PIXIE/PRISM, 
so we can use  the limit found by COBE on $\mu$ and \emph{expected} sensitivities of PIXIE/PRISM 
to evaluate the oscillation probability, $P_{\phi}$. PIXIE/PRISM \cite{Kogut:2011xw,Andre:2013afa} 
will operate in the frequency range 
30 GHz $\leq\nu\leq$ 6 THz or $0.53\leq x\leq 105.9$ and is expect to be more sensitive than COBE. 
PIXIE expects to reach sensitivities  $\mu \sim5\times 10^{-8}$,  $y \sim10^{-8}$, and $\delta T/T \sim10^{-8}$ on a large angular scale, 
which are 3 orders of magnitude better than COBE sensitivity. Here we use the upper bounds on $\mu$ and $\delta T/T$ of COBE\footnote{We remind that COBE did not find spectral distortion of CMB but put only upper values on $\mu$ and $y$.} and the sensitivity of PIXIE which is mostly sensitive to $\mu$-distortion in the energy range $0.53\leq x\leq 11.3$ 
(see Fig. 12 of Ref. \cite{Kogut:2011xw} ).  With the limits on $\mu$ and $\delta T/T$ of a given experiment, 
our goal is to find the minimum value of $g_{\phi\gamma}$ for fixed value of $B$, 
$x_\textrm{min}\leq x\leq x_\textrm{max}$, and $10^{-25}$ eV $\leq m_\phi\leq 10^{-5}$ eV such that both 
Eqs. \eqref{mu} and \eqref{temp-anis} are satisfied.

\begin{figure*}[htbp!]
\centering
\mbox{
\subfloat[\label{fig:Fig5}]{\includegraphics[scale=0.6]{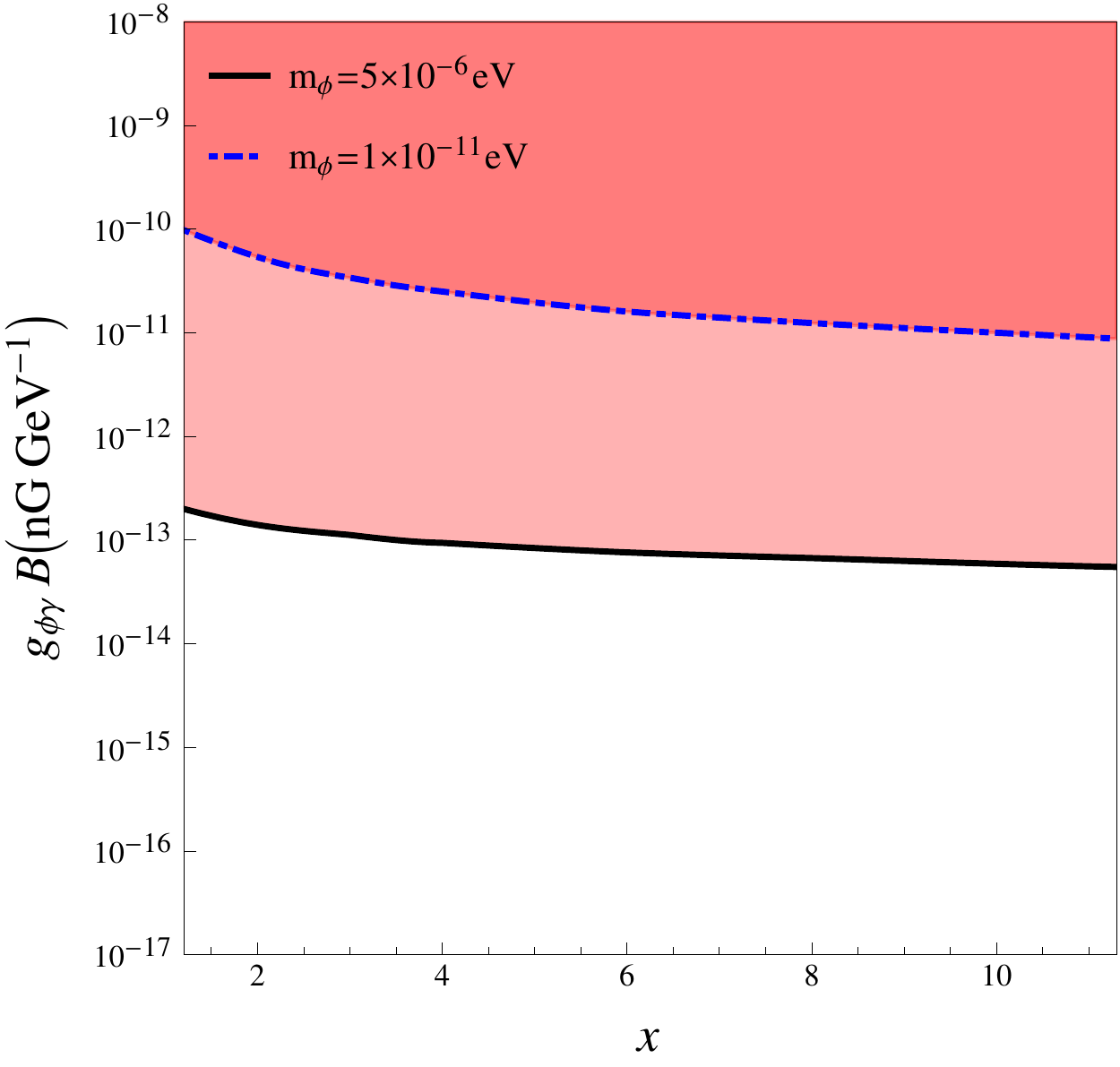}}\qquad
\subfloat[\label{fig:Fig6}]{\includegraphics[scale=0.6]{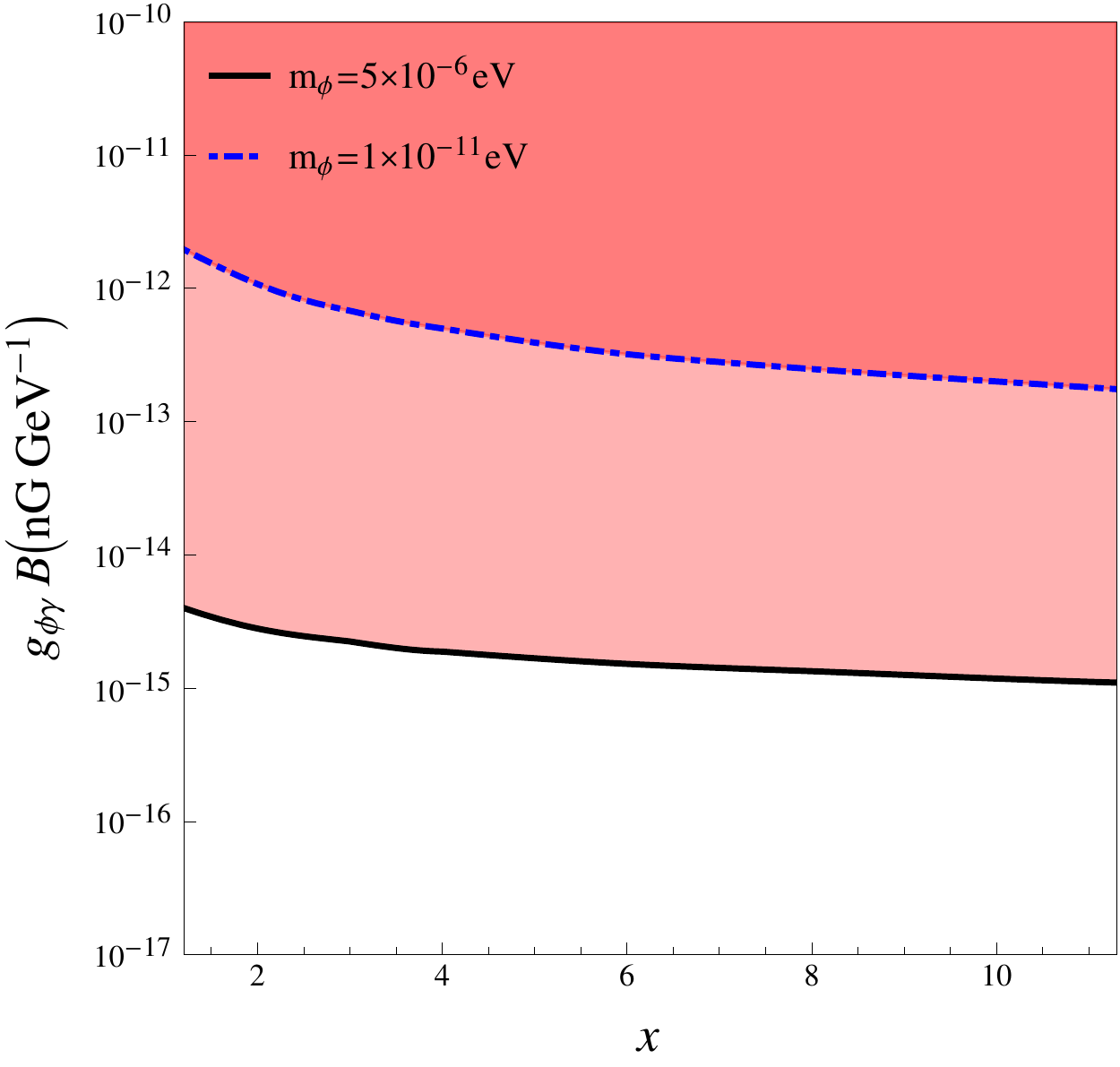}}}
\caption{Exclusion/sensitivity plot for the ALP parameter space  $g_{\phi\gamma} B-x$ due to $\mu$-distortions for magnetic field $B=1$ nG, $m_\phi=1\times 10^{-11}-5\times 10^{-6}$ eV for (a) COBE upper limits on $\mu$ and (b) PIXIE expected sensitivity on $\mu$. We can see that tighter limits on $g_{\phi\gamma} B$ are obtained at the highest energy point, $x=11.3$.}
\label{fig:Fig3a}
\end{figure*}

As a first step in the derivation of these bounds,  
we find which energy channel is most sensitive to $g_{\phi\gamma}B$. Next, after the energy value is
fixed, the second step will be the determination of the allowed area in the $g_{\phi\gamma}-m_\phi$ plane. 
According to Eq. \eqref{prob-alp}, the higher is the photon/ALP energy in the resonant case, the
more stringent, the limits on $g_{\phi\gamma}$ are. However, this kind of behavior is not true in general 
but depends on the energy interval considered. In Fig. \ref{fig:Fig3a} the exclusion and sensitivity plot for the product 
$g_{\phi\gamma}B$ as a function of $x$ for the allowed range of $\mu$-distortion is presented.
In Fig. \ref{fig:Fig5} the exclusion plot based on the COBE data is presented
and in Fig. \ref{fig:Fig6} the sensitivity plot of PIXIE is shown. 
According to  Fig. \ref{fig:Fig5}  tighter bounds on  $g_{\phi\gamma}B$ are obtained for the maximum value of $x=11.3$ independently on the mass, i.e. the resonant one, $m_\phi=5\times 10^{-6}$ eV, or non-resonant, $m_\phi=10^{-11}$ eV. 
We have explicitly verified that this is true for the $\mu$-distortion and temperature anisotropy for every mass 
considered in this paper in the energy range  $0.53\leq x\leq 11.3$. Therefore we focus only on the high energy part of 
the CMB spectrum which is explored by COBE and will be studied by PIXIE, namely, at $x=11.3$ or $\omega_0=11.3\,T_0$. 
For this particular value of $x$, according to Eq. \eqref{mu}, the bound on the oscillation probability  found from the COBE data on the $\mu$-distortion is  
\begin{equation}\label{mu-COBE}
P_{\phi}^{(\mu)}(11.3)\lesssim 9\times 10^{-5}\quad \textrm{(COBE)}.
\end{equation}
The analogous expected sensitivity from PIXIE is:
\begin{equation}\label{mu-PIXIE}
P_{\phi}^{(\mu)}(11.3)\simeq 5\times 10^{-8}\quad \textrm{(PIXIE)}.
\end{equation}
From the bounds on the temperature anisotropy, the more restrictive bound is derived: 
\begin{equation}\label{temp-COBE}
P_{\phi}(11.3)\lesssim 10^{-5}\quad \textrm{(COBE)}
\end{equation}
and the expected sensitivity:
\begin{equation}\label{temp-PIXIE}
P_{\phi}(11.3)\simeq 10^{-8}\quad \textrm{(PIXIE)}.
\end{equation}

Our results have been obtained by solving kinetic equations Eqs. \eqref{rho-gamma-prime} and \eqref{rho-gamma-prime1} 
at two different cosmological epochs. For the case of $\mu$-distortion we have solved Eqs. \eqref{rho-gamma-prime} 
and \eqref{rho-gamma-prime1} in the redshift interval $2.88\times 10^5\leq z\leq 2\times 10^6$ with the initial 
conditions $\rho_\gamma(2\times 10^6)=\rho_{eq}$ and $\rho_\phi(2\times 10^6)=0$ (no ALPs initially). For the case of temperature anisotropy we solved the kinetic equations in the redshift interval $0\leq z\leq 1090$ with the
initial conditions $\rho_\gamma(1090)=\rho_{eq}$ and $\rho_\phi(1090)=0$.

\begin{figure*}[htbp!]
\centering
\mbox{
\subfloat[\label{fig:Fig3}]{\includegraphics[scale=0.6]{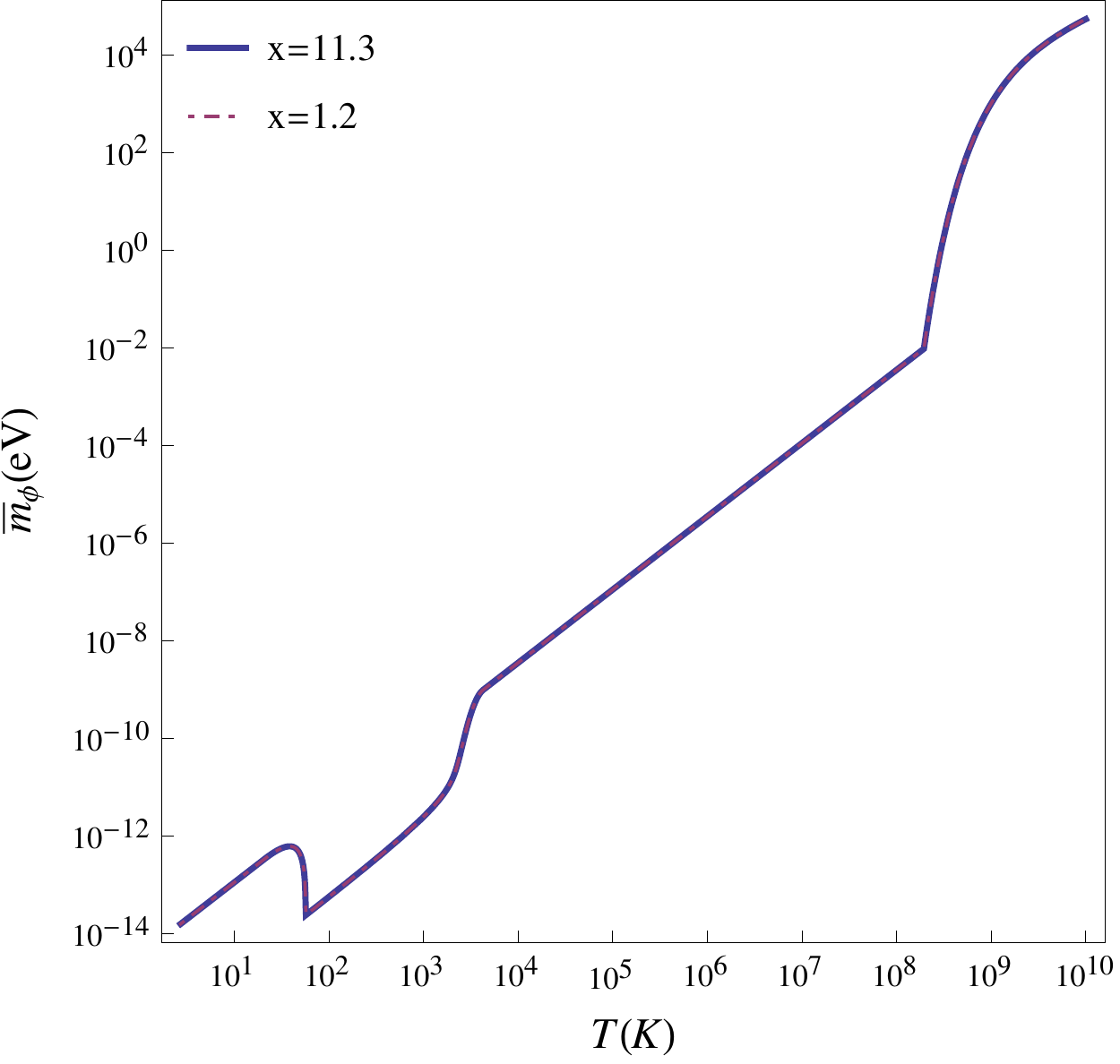}}\qquad
\subfloat[\label{fig:Fig4}]{\includegraphics[scale=0.6]{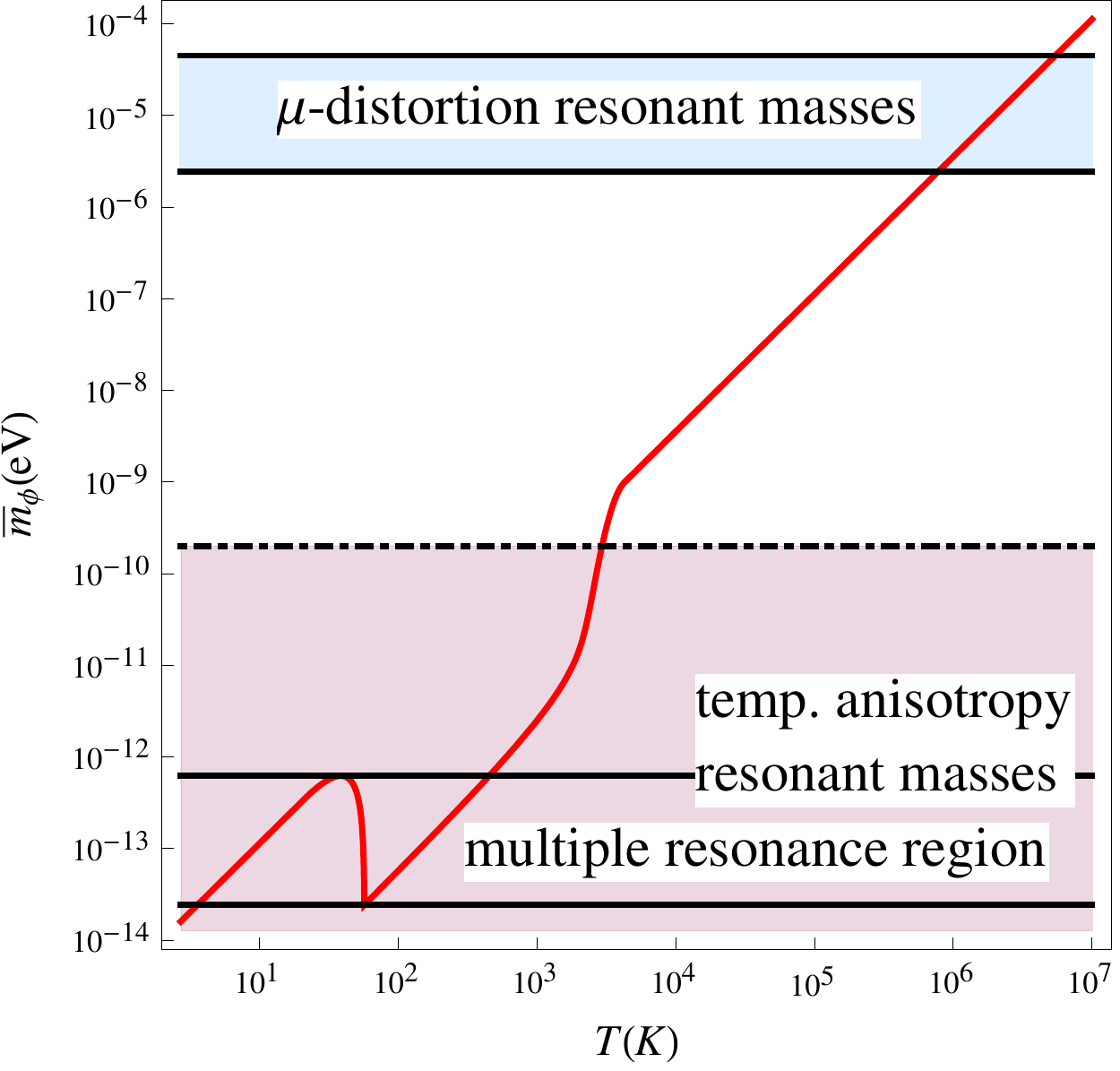}}}
\caption{Resonance mass $\bar m_\phi$ as a function of the temperature $T$ in Kelvin for $B=1$~nG. The resonance mass 
is presented for two values of the photon energy, $x=1.2$ (dot dashed curve) and $x=11.3$ (continuous curve), but the curves
are indistinguishable indicating that the resonance does not depend on $x$ in the energy interval considered here. 
Panel~(b) is the same as panel~(a) but in a reduced temperature interval. 
The resonant mass regions which contribute to $\mu$-distortion and temperature anisotropy, as well as the region where the
multiple resonances are excited (within the solid lines in the lower part of the graph) are presented there.}
\label{fig:Fig4a}
\end{figure*}

In Fig. \ref{fig:Fig4a} the ALP mass as a function of the temperature, for which the resonance transition between photons and ALPs could 
take place in the early universe, is shown. The curves are presented for two values of $x$; $x=1.2$ and $x=11.3$,  but they are indistinguishable. This means that the resonance mass
does not depend on the photon energy, in the range considered here. 
In Fig. \ref{fig:Fig4} the same curves are presented with the enlarged scale and hence in the reduced temperature
interval. The mass values, for which $\mu$-distortion or the temperature anisotropies are induced, are also indicated 
there. The resonant mass range for which $\mu$-distortion is induced, is situated 
from $m_\phi\simeq 4.5\times 10^{-5}$ eV down to $m_\phi\simeq 2.45\times 10^{-6}$ eV. 
The upper extreme corresponds to the redshift $z\simeq 2\times 10^6$ or $T\simeq 5.45\times 10^6$ K, while the 
lower one corresponds to $z\simeq 2.88\times 10^5$ or $T\simeq 7.84\times 10^5$ K. In this paper 
we consider ALPs which are relativistic during the whole epoch $0\leq z\leq 2\times 10^6$, namely,   ALPs with $m_\phi\lesssim T_0$ or $m_\phi\lesssim 10^{-5}$ eV. \footnote{The resonance mass range presented in Fig. \ref{fig:Fig3} much extends the relativistic upper bound, 
$m_\phi\lesssim 10^{-5}$ eV. These calculations are formally done neglecting the fact that the ALPs could be nonrelativistic.
 ALPs with mass $m_{\phi}>10^{-5}$ eV will be considered elsewhere. } 
 In the case of the temperature anisotropy the resonance region extends from 
$m_\phi\simeq 2\times 10^{-10}$ eV ($z\simeq 1090$ or $T\simeq 2970$ K) down to 
$m_\phi\simeq 1.6\times 10^{-14}$ eV at the present epoch ($z=0$). In Fig. \ref{fig:Fig4} the 
region within solid lines labeled ''multiple resonance region'' is the region where multiple resonance transitions occur, 
$2.43\times 10^{-14}$ eV $\leq m_\phi\leq 6.18\times 10^{-13}$ eV.

\begin{table}
\centering
    \begin{tabular}{ | l | l | p{5cm} | }
    \hline
     & Resonant mass range (eV) & Non resonant mass range (eV) \\ \hline
    $\mu$-distortion & $2.45\times 10^{-6}\leq m_\phi\leq 10^{-5}$ & $10^{-25}\leq m_\phi< 2.45\times 10^{-6}$  \\ \hline
    Temperature anisotropy & $1.6\times 10^{-14}\leq m_\phi\leq 2\times 10^{-10}$ & $m_\phi <1.6\times 10^{-14}$ and $ 2\times 10^{-10}< m_\phi\leq 10^{-5}$ \\  \hline      
    Min. value $g_{\phi\gamma}B$(nG$\times$GeV$^{-1}$) \\ for $\mu$-distortion & $5.5 \times 10^{-14}$ (resonant) & $8.76\times 10^{-12}$ (non resonant)\\ \hline
       Min. value $g_{\phi\gamma}B$(nG$\times$GeV$^{-1}$) \\ for temperature anisotropy & $7.2\times 10^{-16}$ (resonant) & $1.9\times 10^{-14}$ (non resonant) \\ \hline
    \end{tabular}
    \caption{Resonance (first column) and non resonance (second column) mass range applied to study the bounds based on $\mu$-distortion (first row) and temperature anisotropy (second row). Minimum value of $g_{\phi\gamma}$ found from the limits on $\mu$-distortion (third row) and on the temperature anisotropy (fourth row).}
    \label{table:1}
\end{table}

\begin{figure*}[htbp!]
\centering
\mbox{
\subfloat[\label{fig:Fig7}]{\includegraphics[scale=0.56]{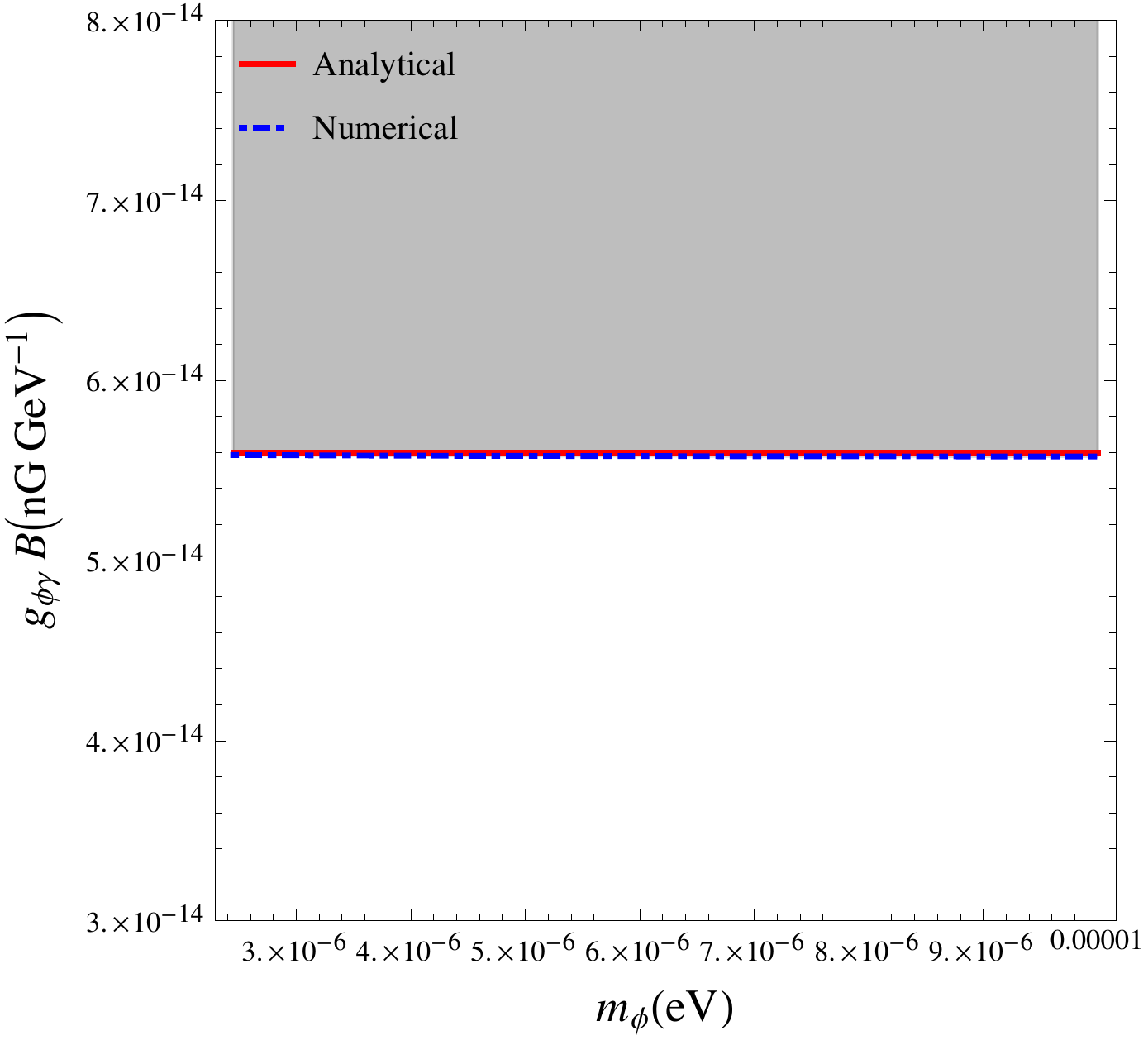}}\qquad
\subfloat[\label{fig:Fig8}]{\includegraphics[scale=0.6]{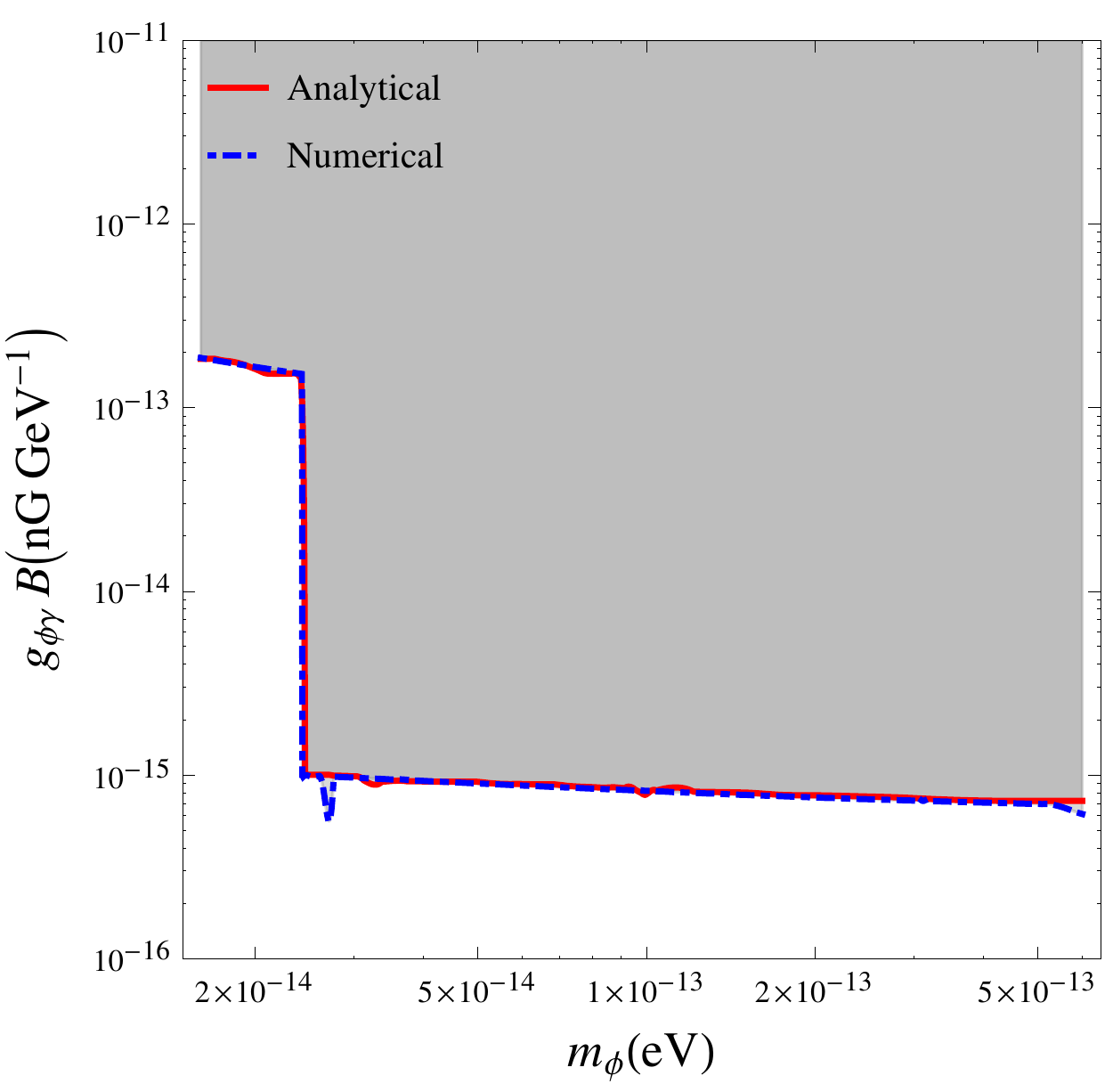}}}
\caption{Exclusion plot based on the COBE data for the ALP parameter space, $g_{\phi\gamma}B-m_\phi$, 
for magnetic field $B= 1$~nG and $x=11.3$ in the resonant case. In panel~(a) the exclusion plot is deduced from upper limit on $\mu$ and in panel~(b) the 
exclusion plot is derived from the COBE sensitivity on the temperature anisotropy $\delta T/T\sim 10^{-5}$ . In panel~(b) the bounds are presented for the mass in the resonant mass range: 
$1.6\times 10^{-14}$ eV $\leq m_\phi\leq 6.18\times 10^{-13}$ eV.}
\label{fig:Fig5a}
\end{figure*}

In Fig. \ref{fig:Fig5a} the exclusion area in the $g_{\phi\gamma}B-m_\phi$ plane, derived from numerical and analytical solutions of the kinetic equation and the COBE limits on $\mu$ and $\delta T/T$, is presented for the resonant ALP masses. In Fig. \ref{fig:Fig7},  a comparison between the 
analytical and numerical results is presented for the case of $\mu$-distortion and in Fig. \ref{fig:Fig8} the same is done
for  the temperature anisotropy. The limits based on the analytic solution have been obtained using Eqs. \eqref{res-T} and \eqref{rho-res1}. 
Temperature anisotropy limits have been derived in the case when QED effects are negligible in comparison with  those induced by $m_a$. In the energy range 
considered in this paper Eq. \eqref{rho-res1} is not only valid at the post recombination epoch but also 
during $\mu$ epoch, namely the epoch corresponding to redshift $2.88\times 10^5\lesssim z\lesssim2\times 10^6$.
In Fig. \ref{fig:Fig7},  a perfect agreement between analytical and numerical results, found by solving the kinetic equation in the $\mu$ 
epoch, is observed, where the absolute difference between the two methods is  less than $10^{-2}$. 

In Fig. \ref{fig:Fig8} a
comparison between analytical and numerical results is presented for the case of the temperature anisotropy in 
the ALP mass range $1.6\times 10^{-14}$ eV $\leq m_\phi\lesssim 6.18\times 10^{-13}$~eV. 
It is only in this mass range that there are multiple resonance transitions. We see that there is a good agreement between numerical and analytical results. 
Analytical calculations demonstrate that the first resonance transition gives stronger limits on $g_{\phi\gamma}B$ for fixed $\bar m_\phi$. 
So, in the analytical treatment we considered only the ALP masses which cross the resonance in the temperature interval 
$57.24$ K $\lesssim T\lesssim 450$ K. In this temperature range, the ionization fraction is below unity. The later resonance 
transitions for the same ALP masses give weaker limits on $g_{\phi\gamma}B$ but have the same impact on the temperature anisotropy.

\begin{figure*}[h!]
\centering
\includegraphics[width=15cm, height=10cm]{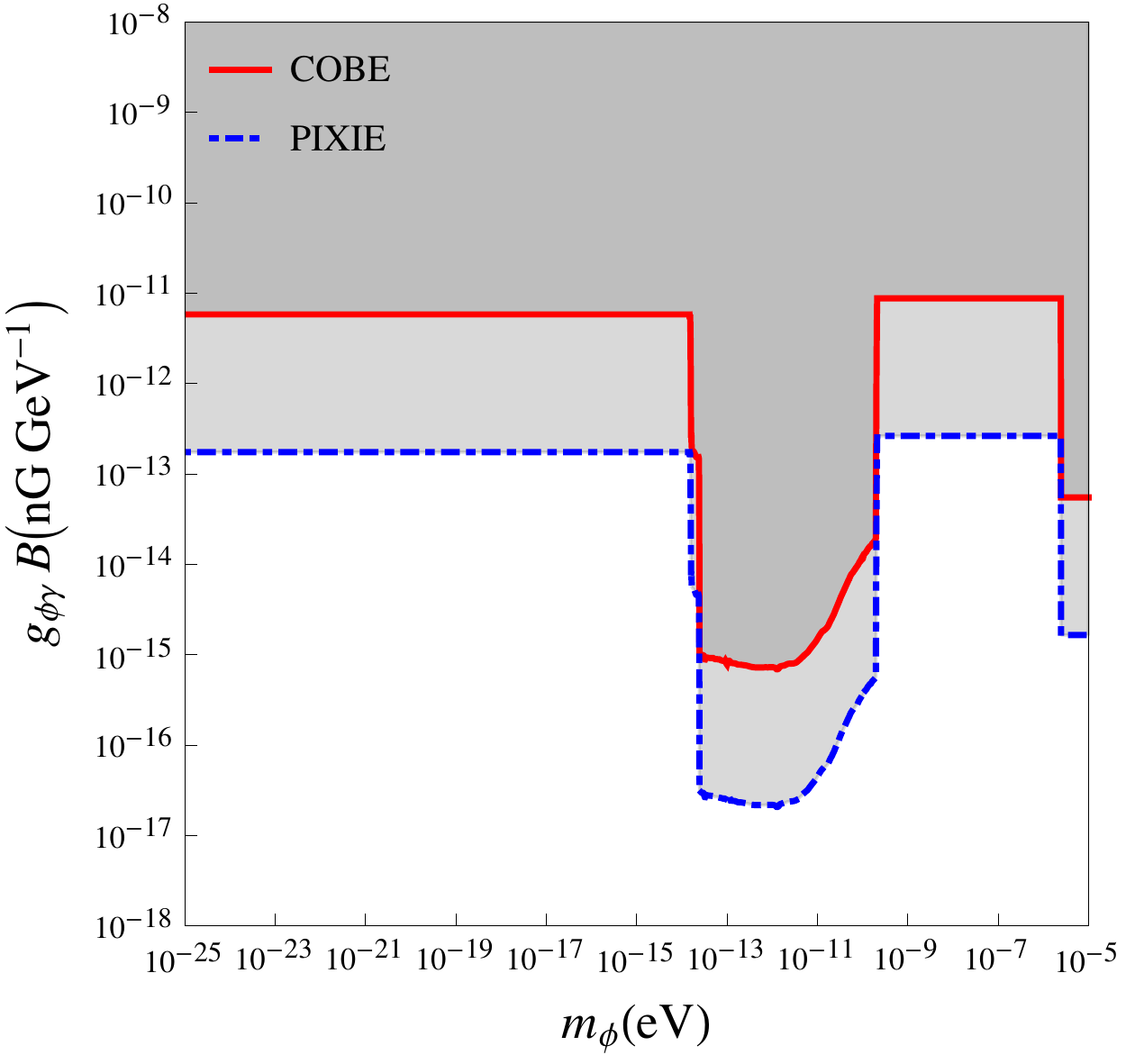}
\caption{Exclusion plot for the ALP parameter space, 
$(g_{\phi\gamma}B-m_\phi)$ in the case of COBE (region above solid line) and sensitivity of PIXIE (region above dot dashed line), based on the
combined limits on $\mu$-distortion and temperature anisotropy for magnetic field $B=1$ nG and $x=11.3$. We can see that at the transition points ($m_\phi=1.6\times 10^{-14}$ eV, $m_\phi=2\times 10^{-10}$ eV, and $2.45\times 10^{-6}$ eV) from the resonant region to the non resonance region there is an extremely sharp increase in $g_{\phi\gamma}B$, 
by more than an order of magnitude. This is a consequence of the fast varying solutions at the transition points. 
Using the logarithmic scale and wide range of ALP masses slightly magnifies this effect. In the ALP mass range $m_\phi\leq 2\times 10^{-10}$ eV, the exclusion plot of COBE and sensitivity of PIXIE,  have been obtained from CMB temperature anisotropy. For $2\times 10^{-10}<m_\phi\leq 10^{-5}$ both the exclusion plot and sensitivity of PIXIE have been obtained from $\mu$ distortion. In the mass range $m_\phi\leq 2\times 10^{-10}$ eV, temperature anisotropy give more stringent limits with respect to $\mu$-distortion while for $m_\phi>2\times 10^{-10}$ eV, $\mu$-distortion gives tighter limits in comparison with temperature anisotropy.}
\label{fig:Fig9}
\end{figure*}

In Fig. \ref{fig:Fig9} the exclusion plot based on the  COBE data and a possible ALP discovery potential, based on the  
expected sensitivity by PIXIE, are shown. Our results are obtained from the 
numerical solution of the kinetic equation for the  combined $\mu$-distortion and 
temperature anisotropy. The exclusion (sensitivity) limits, based on  the COBE (PIXIE, future) measurements  
in the ALP mass range  $2\times 10^{-10}$ eV $< m_\phi\leq 10^{-5}$ eV, have been obtained from the CMB $\mu$ distortion.
For the mass range  $m_\phi\leq 2\times 10^{-10}$ eV, the region above the solid line is excluded by the observed CMB temperature anisotropy, 
so no improvement by PIXIE is expected here. Returning to $\mu$-distortion, we note that 
for $2\times 10^{-10}$ eV $< m_\phi\leq 10^{-5}$ eV the region above the solid line is excluded due to the nonobservation of CMB $\mu$-distortion at the level of $\mu\leq 9\times 10^{-5}$. In the case of PIXIE our limits are presented as expected sensitivity and the region above the dot-dashed curve is the region, in which PIXIE could find, in principle lower temperature anisotropy and chemical potential with respect to COBE. At the level of temperature anisotropy $\delta T/T\sim 10^{-8}$, PIXIE can, in principle, detect temperature anisotropy coming from the transition of CMB photons into ALPs.

In both cases (COBE and PIXIE), stronger limits on $g_{\phi\gamma}B$ are obtained in the resonant mass region 
(see i.e. Fig. \ref{fig:Fig4} and Tab. \ref{table:1}) and weaker limits are obtained in the non-resonant case. 
An extremely rapid increase of the upper bound on the product $g_{\phi\gamma}B$ 
in the transition from resonance to non-resonant mass region can be explained by a strong increase of the resonance probability in comparison with the non-resonant one.

We can see from Fig. \ref{fig:Fig9} that  PIXIE can confine the product $g_{\phi\gamma}B$   better than COBE
roughly by the factor
 $\sqrt{\mu_\textrm{PIXIE}/\mu_\textrm{COBE}}$ for the $\mu$-distortion and 
 $\sqrt{(\delta T/T)_\textrm{PIXIE}/(\delta T/T)_\textrm{COBE}}$ for the temperature anisotropy. 

\section{\label{sec:6}Discussion and conclusion}

We have derived the constraints on the ALP parameters, in particular, on the product  $g_{\phi\gamma}B$,
using the COBE upper bound on $\mu$ and sensitivity on $\delta T/T$. With the expected sensitivity of PIXIE the constraints would be strongly improved
or the spectral distortion and possibly very small temperature fluctuations could be found.
The constraints based on the COBE data update and further improve earlier ones, which have been obtained only in the resonant case; 
see Ref. \cite{Mirizzi:2009nq}. Moreover, we have also found  the bounds on the parameters in the non resonant case which was not 
considered in the literature. Our results have been obtained by taking into account the coherence breaking of photons in the cosmological 
plasma due to non-forward scattering and absorption. To find the conversion probability, the  quantum kinetic equation 
was solved numerically and analytically  in the steady state approximation.

The impact of photon to ALP transition on CMB depends on the magnitude of the large scale cosmological magnetic field, $B$,
which is rather poorly known. Mostly only upper limits on $B$ are established. So we cannot put limits on
the coupling constant, $g_{\phi\gamma}$, or ALP mass but only on the product $g_{\phi\gamma}B$ as a function of 
the ALP mass, as it is presented in  Fig. \ref{fig:Fig9}. We have done the analysis in the ALP mass interval
 $10^{-25}$ eV $\leq m_\phi\leq 10^{-5}$ eV, which is especially interesting, since for most masses in this interval the
 resonance transition at the post BBN epoch could be induced, and sometimes even more than once. 
 As we can see from Fig. \ref{fig:Fig4a}, the range of masses which leads to the resonant transition at different temperatures is very wide starting from  $m_{\phi}=1.6\times 10^{-14}$ eV (the resonance transition at the present time) up 
 to $m_\phi= 10^{-5}$ eV (the maximum mass considered in this paper to ensure that ALPs are relativistic).
We have also considered the mass ranges at which no resonant transition takes place at the post BBN epoch, 
namely, we study the masses in the interval $10^{-25}$ eV $\leq m_{\phi}\leq 1.6\times 10^{-14}$ eV. 

We would like to mention two important differences with other works, where these phenomena were also considered. 
First, there is a disagreement with Ref.~\cite{Mirizzi:2009nq} where it was claimed that the ALP mass range, which causes resonant 
transition at the post BBN epoch, extends roughly speaking from  $m_{\phi}=10^{-14}$ eV up to $m_\phi=10^{-2}$ eV, 
while in our case the resonance mass goes up to $m_\phi  \sim 10^4$~eV, see Eq. \ref{m-times} and 
Fig.~\ref{fig:Fig3}. 
This difference appears because in Ref. \cite{Mirizzi:2009nq} the increase (going backward in time) of the free electron 
density prior the $e^+e^-$-annihilation epoch was not apparently taken into account. 

The second difference is found with Ref.~\cite{Tashiro:2013yea}, 
where the photon-ALP resonance transition has been extended down to the ALP masses $m_\phi\lesssim 10^{-14}$ eV, 
while according to our calculations 
the resonance is excited (up to the present time) only for the masses above this value. This result agrees 
with that of Ref.~\cite{Mirizzi:2009nq}. The bounds on $g_{\phi\gamma}B$ (based on the COBE data), which are 
derived here, in some part of the ALP parameter space, confirm those obtained by other methods, while in other parts of ALP parameter space we presented new results. In particular, as one can see from Fig.~\ref{fig:Fig9}, for the ALP mass $ m_{\phi}\lesssim 10^{-10}$ eV our result is 
$g_{\phi\gamma}B\lesssim 1.2\times 10^{-14}\, \textrm{nG}\times \textrm{GeV}^{-1}$ (for COBE limits) which is 3 
orders of magnitude stronger than that presented in Ref.~\cite{Brockway:1996yr}: 
$g_{\phi\gamma}B \lesssim 10^{-11}\, \textrm{nG}\times \textrm{GeV}^{-1}$ for 
$m_\phi\lesssim 10^{-10}$ eV.  
  
We would like to emphasize an importance of the numerical solution of the quantum kinetic equation in obtaining the bounds on the ALP 
parameter space. We also would like to mention that the analytical estimate of the resonance transition probability according to 
Eq.~\eqref{prob-alp} is in perfect 
agreement, with the numerical results. 
We observe that at the post recombination epoch multiple resonance transitions occur in the temperature interval 
$4.1$~K~$\lesssim T\lesssim 450$~K. However, the first resonance transition occurring in the temperature interval $57.24$~K~$\lesssim T\lesssim 450$~K leads to stronger limits on $g_{\phi\gamma}B$. 
Therefore careful attention must be paid to using Eq. \eqref{rho-res1} to set limits on 
the ALP parameter space.
If Eq. \eqref{rho-res1} is used to evaluate the transition probability for temperature $T\lesssim 57.24$~K, one would erroneously find weaker limits for $g_{\phi\gamma}B$ in comparison with limits found in temperature interval $57.24$ K $\lesssim T\lesssim 450$ K. This is due to the fact that for  $T\lesssim 57.24$~K the second and the third resonance transitions occur, which lead to weaker limits for $g_{\phi\gamma}B$ than  the
first resonance transition. 
We would like to stress that in order to get reliable numerical results very high machine working precision
and accuracy are required. If normal machine precision and accuracy are used, one would erroneously find much weaker 
results on the ALP parameter space. 

Let us recall that the bounds presented here are expressed in the form of 
the product $g_{\phi\gamma}B$.
So better information about the magnitude of the cosmological magnetic field is desirable.
It would allow us to derive more accurate bound on the coupling constant. Higher values of the magnetic field 
would amplify the oscillation probability and consequently the CMB restrictions on the ALP parameter 
space would be more stringent. 

Let us note that we solved the kinetic equations assuming that no ALPs were present initially. 
However, the validity of this assumption depends on the ALP production mechanism, and it is not excluded that the initial abundance of ALPs was non negligible. Indeed, they can be produced for example by the misalignment mechanism in the early 
universe, as it happened with axions. However,
if ALPs were produced by the misalignment mechanism, they would be concentrated in a different energy range
than CMB, see Ref. \cite{Cadamuro:2011fd}. 
Some ALPs could be produced by the photon transformation into them, so they would have the same energy as the CMB photons 
but the density of such ALPs would be very small. Strictly speaking we cannot exclude that there is some production mechanism 
of ALPs in the same energy range 
as the CMB photons. In this case the limits on the ALP parameters would be weakened. 

In this work we considered only $\mu$ and temperature anisotropy limits on the
ALP parameter space and did not consider $y$ limits. Indeed, in order 
to set limits on the ALP parameter space based on $y$-distortion
it is necessary to have an analytical solution for the photon distribution $n_\gamma$. However, as we have already discussed, it is very difficult 
to find an analytic solutions and without an analytic expression for $n_\gamma$, we are unable to derive limits which follow from the analysis of $y$-distortion. We anticipate that if one could use the $y$ parameter to set limits on ALP parameter space, it would be possible to obtain very stringent limits on $g_{\phi\gamma}B$ especially for those ALP masses which hit the resonance during the $y$ epoch.

The bounds and expected sensitivity derived in our work are based on the solution of the linearized form of the quantum kinetic equation, \eqref{eq-dens-matrix} in the steady state approximation. The corresponding
system of Eq.~\eqref{densitysys} is highly stiff and is very difficult to solve numerically. However, we have 
checked that in the cases when the solutions are stable they, well agree with those found in the steady state 
approximation. Moreover, we expect to use the quantum kinetic equations beyond 
the linearized approximation and to obtain more rigorous bounds on the ALP parameter space. 

At the end we would like to emphasize that kinetic equations for the density operator are most appropriate to study the phenomena considered in this paper. In general, when the oscillation length of the photon-axion system is much
 longer that the mean free path of photons or axions in the plasma, the usage of the density matrix is mandatory to take into account the effects of coherence breaking. The wave function formalism is not proper in this case. We have found that in the resonant case, the density matrix formalism gives stronger limits on $g_{\phi\gamma}B$ by more than an order of magnitude in comparison with the wave function approximation. However, in the wave function approximation, only the resonant case has been considered in the literature, while the non resonant case has not been studied at all. Therefore we cannot compare the density matrix results, in the non resonant case, with the wave function approximation.

 {\bf{Acknowledgments}} 
: We acknowledge the support of the Russian Federation
government, Grant No. 11.G34.31.0047. D. E. thanks the hospitality of the Novosibirsk State University where this manuscript was initiated.

  \end{document}